






\magnification\magstep1
\baselineskip14pt
\vsize23.5truecm 

\overfullrule=0pt 




\def\hatt{\widehat}
\def\dell{\partial}
\def\tilda{\widetilde}
\def\eps{\varepsilon}

\def\half{\hbox{$1\over2$}}

\def\arr{\rightarrow}
\def\normal{{\cal N}}

\def\RR{\mathord{I\kern-.3em R}}
\def\PP{\mathord{I\kern-.3em P}}
\def\NN{\mathord{I\kern-.3em N}}
\def\ZZ{\mathord{I\kern-.3em Z}} 
\def\Var{{\rm Var}}

\def\d{{\rm d}}

\def\mtrix{\pmatrix} 

\def\subsection{\medskip}
\def\tr{{\rm t}}

\font\bigbf=cmbx12

\font\csc=cmcsc10

 at 10truept 
\font\smallrm=cmr8

\def\today{\number\day \space \ifcase\month\or
January\or February\or March\or April\or May\or June\or 
July\or August\or September\or October\or November\or December\fi  
\space \number\year}


   
\def\ref#1{{\noindent\hangafter=1\hangindent=20pt
  #1\smallskip}}          

\def\quotationone{\smallrm Where there is a Will}
\def\quotationtwo{\smallrm There is a Won't}
\def\hskipdistanceleft{\hskip-3.5pt}
\def\hskipdistanceright{\hskip-2.0pt}
\footline={{
\ifodd\count0
        {\hskipdistanceleft\quotationone\phantom{\smallrm\today}
                \hfil{\rm\the\pageno}\hfil
         \phantom{\quotationone}{\smallrm\today}\hskipdistanceright}
        \else 
        {\hskipdistanceleft\quotationtwo\phantom{\today}
                \hfil{\rm\the\pageno}\hfil
         \phantom{\quotationtwo}{\smallrm\today}\hskipdistanceright}
        \fi}}

         
\def\cstok#1{\leavevmode\thinspace\hbox{\vrule\vtop{\vbox{\hrule\kern1pt
        \hbox{\vphantom{\tt/}\thinspace{\tt#1}\thinspace}}
        \kern1pt\hrule}\vrule}\thinspace} 
 


\def\fermat#1{\setbox0=\vtop{\hsize4.00pc
        \smallrm\raggedright\noindent\baselineskip9pt
        \rightskip=0.5pc plus 1.5pc #1}\leavevmode
        \vadjust{\dimen0=\dp0
        \kern-\ht0\hbox{\kern-4.00pc\box0}\kern-\dimen0}}


\def\sumjn{\sum_{j=1}^n}
\def\Splus{{\tt S-Plus}}

\input miniltx
\input graphicx


\def\today{May 1993}
\def\quotationone{\smallrm Hjort and Lumley}
\def\quotationtwo{\smallrm NLH plots}

\centerline{\bigbf Normalised Local Hazard Plots}

\smallskip
\centerline{\bf Nils Lid Hjort$^{a,b}$ and Thomas Lumley$^b$}

\smallskip
\centerline{\sl University of Oslo$^a$ and University of Oxford$^b$}

\smallskip
{{\smallskip\narrower\baselineskip12pt\noindent
{\csc Abstract}. 
The purpose of this paper is to develop 
and illustrate certain classes of graphical plots that can be 
used for model verification in quite general survival data 
and life history data models. 
By suitably comparing nonparametric and parametric estimates
of hazard rate functions over time a hazard comparison function 
can be constructed which under parametric model assumptions
is approximately a zero-mean normal process. 
The test curves we propose are locally normalised versions
of such hazard comparison functions. 
Under model conditions the test function is approximately 
a standard normal for each time point. 
This makes the normalised local hazard curves easy to interpret.
We give explicit constructions for the most commonly used models
of survival analysis, including the exponential, the Weibull,
the Gompertz, the gamma, and for parametric Cox regression. 
Algorithms carrying this out have been developed in \Splus. 
Various theoretical and practical issues are discussed, including
detection power and extensions to time-discrete models. 
Illustrations are given on simulated and real data.

\smallskip\noindent
{\csc Key words:} \sl 
counting processes, 
goodness of fit, 
hazard function models, 
normalised local hazard plots,
parametric Cox model,
\Splus, 
time-discrete hazard models
\smallskip}}

\bigskip
{\bf 1. Introduction and summary.} 
This paper discusses and illustrates certain
graphical plots that can be used for model verification purposes 
in survival data situations and in fact also in much more general 
counting process models for life history data. 
Such models involve certain hazard rate functions. 
By suitably comparing nonparametric and parametric estimates
of hazard rate functions over time we are able to produce a 
hazard comparison function which under the conditions of the 
parametric model is approximately a zero-mean normal process. 
And by normalising this local hazard function with a local
estimate of the standard deviation we end up with a test function 
which is approximately a standard normal for each time point,
provided the parametric model studied is valid. 

We call these test curves {\it normalised local hazard plots}. 
Such a test curve is a concise summary of the discrepancy 
between the parametric and nonparametric model, and 
is easy to judge by eye, since the precision is the same throughout
the time axis, and since everybody knows the standard normal. 
If the model is right then the test curve should stay inside 
the $\pm1.96$ horizontal band most of the time, 
and incorrect aspects of the model will 
lead to curves that wander outside this band. 

There are also formal goodness of fit testing procedures,  
see Hjort (1990) and Andersen, Borgan, Gill \& Keiding (1993, chapter VI),
the latter henceforth referred to as ABGK. 
Such tests are valuable since they can guarantee e.g.~a maximum 
5\% chance of incorrectly rejecting a model, of course, 
but the emphasis here is on graphical checking tools for various models. 
The plots will be useful when the statistician explores 
different modelling opportunities and can be more informative than
the simple yes or no answer that a formal test provides, 
in that one is shown in what way or ways a given model does not hold. 
The formal tests can in this light be regarded as giving 
supplementary confirmatory information. 

To construct the test curves properly one essentially needs 
(a) proofs of limiting normality of the processes,
(b) expressions for and consistent estimates for the variances of
the limit processes. Sufficient theory for this 
has in fact already been developed in Hjort (1985, 1990). 
The model test plot idea was also briefly mentioned 
in these papers, as an attractive consequence of the general results.
Arulchelvam (1992) implemented one version of these plots 
for a couple of models, and she illustrated their use 
on simulated and real data. 
The intention of the present paper is to systematically work out 
variance estimation and other necessary details for the
most popular models, and to demonstrate the usefulness 
of the plots on real and simulated data. 

For concreteness we choose to develop and discuss the plots 
mainly in the context of `traditional survival analysis',
involving possibly right-censored observations of life-times,
with or without covariate information. The basic methods and 
results hold with suitable modifications for general counting process
models for life history data, as briefly discussed in section 6. 
The traditional framework for homogeneous data is as follows: 
$T_1^0,\ldots,T_n^0$ are i.i.d.~life-times from a distribution
with hazard rate $h(s)$ and cumulative hazard rate $H(t)=\int_0^t h(s)\,\d s$.
The data may be censored from the right,
so what one observes is $t_j=\min(t_j^0,c_j)$ and $\delta_j=I\{t_j^0\le c_j\}$,
where $c_j$ is the possibly interfering censoring time. 
The censoring mechanism is assumed noninformative 
(see ABGK for discussion of this). 
A parametric model is considered, say of the form $h(s)=h(s,\theta)$ 
for some unknown parameter $\theta=(\theta_1,\ldots,\theta_p)^\tr $. 

It is convenient to define and study both parametric 
and nonparametric estimation in terms of 
$$N(t)=\sumjn I\{t_j\le t,\delta_j=1\}  
	\quad {\rm and} \quad
	Y(t)=\sumjn I\{t_j\ge t\}. \eqno(1.1)$$
Here $N(t)$ is the counting process, counting the number of observed 
failures, while the left-continuous $Y(t)$ is the at risk process, 
counting the number among the $n$ original items that are still 
at risk just prior to time point $t$. Since $N(.)$ is flat between
observed life-times $\int_0^tg(s)\,\d N(s)$ means simply 
$\sum_{j\colon t_j\le t}g(t_j)\delta_j$. Nonparametric estimation of 
$H$ is carried out using the Nelson--Aalen estimator 
$$\hatt H(t)=\int_0^t\d N(s)/Y(s)=\sum_{j\colon t_j\le t}\delta_j/Y(t_j). 
	\eqno(1.2)$$
%
As for parametric estimation, we work with the maximum likelihood
(ML) estimator $\hatt\theta$. The log-likelihood can be written 
$\int_0^\tau\{\log h(s,\theta)\,\d N(s)-Y(s)h(s,\theta)\,\d s\}$, 
where $[0,\tau]$ is the finite or infinite time interval over 
which the processes are observed,
see for example ABGK (chapter VI). 

Hjort (1990) studied a large class of goodness of fit processes of the type
$$D_n(t)=\sqrt{n}\int_0^tK_n(s)\{\d\hatt H(s)-h(s,\hatt\theta)\,\d s\}
=\sqrt{n}\int_0^t{K_n(s)\over Y(s)}\{\d N(s)-Y(s)h(s,\hatt\theta)\,\d s\}.
	\eqno(1.3)$$
This local hazard comparison function should take values around zero 
if the model holds. The weight function $K_n(s)$ 
can be chosen to suit different aims, as exemplified later, 
and is here meant to be scaled in a stable way, so that it has a 
well-defined limit in probability function $k(s)$ as $n$ grows. 
Special cases of interest include the following,
to be referred to later as Type A, Type B, Type C: 
$$\matrix{
D_n(t)&=\sqrt{n}\{\hatt H(t)-H(t,\hatt\theta)\}, \hfill 
	&{\rm for\ }K_n(s)=1, \hfill \cr
D_n(t)&=n^{-1/2}\{N(t)-\int_0^t Y(s)h(s,\hatt\theta)\,\d s\}, \hfill 
	&{\rm for\ }K_n(s)=Y(s)/n, \hfill \cr
D_n(t)&=n^{-1/2}\int_0^t G_n(s)\{\d N(s)-Y(s)h(s,\hatt\theta)\,\d s\}, \hfill
	&{\rm for\ }K_n(s)=\{Y(s)/n\}G_n(s). \hfill \cr}
	\eqno(1.4)$$	
Type A uses $K_n(s)=1$, and then $D_n(.)$ 
directly compares nonparametric versus parametric estimates of 
the cumulative hazard function.
Type B employs $K_n(s)=Y(s)/n$, the proportion at risk at time $s$,
in which case $D_n(.)$ compares the observed number of failures 
with the predicted number of failures under the parametric model. 
There are other ways of predicting the number of failures 
in the course of a time interval from the model, 
but this is a natural way based directly on the hazard rate interpretation;
$\d N(s)$, the number of failures in $[s,s+\d s]$ among the $Y(s)$
that are observed to be at risk just prior to time $s$, 
is simply a binomial with probability parameter $h(s,\theta)\,\d s$. 
This leads to $\int_a^bY(s)h(s,\hatt\theta)\,\d s$ as the `expected'
version of $N[a,b]$. 
Finally Type C, with $K_n(s)=\{Y(s)/n\}\,G_n(s)$,
is as general as (1.3), of course, but some interesting procedures
that are tailor-made to have optimal detection power against certain
departures from the model are of this form, sometimes with 
a deterministic weight function $G_n(s)=g(s)$, as exemplified in 3.2. 

Under a mild set of regularity conditions 
$D_n(.)$ tends in distribution to a certain zero-mean Gau\ss ian
process $D(.)$, see section 2. The basic {normalised local hazard plot} 
idea, as explained in the introductory paragraphs, is to draw the curve 
$${\rm NLH}(t)=D_n(t)/\hatt\kappa(t) \quad 
	{\rm versus\ time\ }t\in(0,\tau), \eqno(1.5)$$
where the denominator is an estimate of the local standard deviation. 
This curve has the property of being approximately a standard
normal {\it for each $t$}, provided the model is correct. 

The rest of the paper is organised as follows:
Section 2 discusses some options for $\kappa(t)$ estimation and 
gives detailed general descriptions for three types of NLH-plots.
This is next applied in section 3 to a selection of the most 
popular parametric models of survival analysis: 
The exponential model with a constant hazard rate, 
the Weibull, the gamma, et cetera. 
In section 4 similar goodness of fit plots are developed and illustrated
for the case of a parametric Cox regression model. 
Behaviour of the plots outside model conditions,
and the plots' detection power against various kinds of alternatives,
is discussed in section 5. Section 6 shows how the previous methods and 
results extend to more general models for life history data, like
time-inhomogeneous Markov chains. Although our emphasis is on
time-continuous models and methods we take time out in section 7
to present the essential formulae for time-discrete hazard rate models. 
A number of supplementary remarks, including comparisons with 
other graphical goodness of fit methods that have been proposed
in the literature, are placed in section 8. 

The paper concludes in visual mode, providing a number of 
illustrations on real and simulated data in section 9. 
User-friendly well-commented algorithms producing the different plots,
as well as the necessary parameter estimates and standard deviations, 
have been implemented in \Splus. These are available from the 
authors upon polite request 
and are briefly described in the Appendix. 

\bigskip
{\bf 2. The three basic types of NLH-plots.}
We need the limit distribution of $D_n(t)$ of (1.3), and 
we need consistent estimates of the limit variance function.

\subsection
{\csc 2.1. Limit distribution of the goodness of fit process.}
Some further notation and results are needed at this stage.
A basic ingredient in the large-sample analysis of the
behaviour of both parametric and nonparametric estimators 
is the limiting process $V(t)$ of the martingale 
$n^{-1/2}\{N(t)-\int_0^tY(s)h(s)\,\d s\}$. This 
is a zero-mean Gau\ss ian process with independent
increments of size $\Var\{\d V(s)\}=y(s)h(s)\,\d s$, where  
$y(s)$ is the limit in probability function of $Y(s)/n$, 
i.e.~the limiting proportion of items at risk at time $s$. 
One can show that $\sqrt{n}\{\hatt H(t)-H(t)\}$ 
tends in distribution to $\int_0^ty(s)^{-1}\,\d V(s)$,
see ABGK (chapter IV). 
%
Let next $\psi(s,\theta)={\dell\over \dell\theta}\log h(s,\theta)$, 
a $(p\times1)$-vector. The $p\times p$-matrix
$$\Sigma=\int_0^\tau \psi(s,\theta)\psi(s,\theta)^\tr y(s)\,h(s,\theta)\,\d s
	\eqno(2.1)$$
will enter several of our calculations. 	
The ML estimator solves 
$\int_0^\tau\psi(s,\theta)\{\d N(s)-Y(s)\allowbreak h(s,\theta)\,\d s\}=0$. 
It is consistent and satisfies
$$\sqrt{n}(\hatt\theta-\theta)\arr_d\Sigma^{-1}
	\int_0^\tau\psi(s,\theta)\,\d V(s)
	\sim\normal_p\{0,\Sigma^{-1}\}. \eqno(2.2)$$

The limit of $D_n(t)$ can now be described: 
Under natural and mild regularity conditions, which include 
convergence in probability of the weight function $K_n(s)$ 
to an appropriate $k(s)$, it converges in distribution to 
$$D(t)=\int_0^t\{k(s)/y(s)\}\,\d V(s)
-\Bigl(\int_0^tk(s)h(s,\theta)\psi(s,\theta)\,\d s\Bigr)^\tr \Sigma^{-1}
 \int_0^\tau\psi(s,\theta)\,\d V(s). $$
This is proved in Hjort (1990, section 2). 
The limit is a zero-mean Gau\ss ian process,
and some calculations show that the variance is 
$$\kappa(t)^2=\int_0^t{k(s)^2\over y(s)}\,h(s,\theta)\,\d s
-\Bigl(\int_0^tk(s)h(s,\theta)\psi(s,\theta)\,\d s\Bigr)^\tr 
\Sigma^{-1}\Bigl(\int_0^tk(s)h(s,\theta)\psi(s,\theta)\,\d s\Bigr). \eqno(2.3)$$
Of course the covariance needs to be given too in order to
fully describe the $D(.)$ process, but our primary interest 
in this paper lies with the NLH-plots (1.5). 
Therefore the variance $\kappa(t)^2$ is the quantity we need to estimate,
at least for $k(.)$ functions corresponding to Type A, Type B, Type C
encountered in (1.4). 

\subsection
{\csc 2.2. Estimating the variance.}
There are several choices, each with its own merits. 
A parametric plug-in option is to use 
$h(s,\hatt\theta)$ and $\psi(s,\hatt\theta)$, together with 
$\hatt y(s)=Y(s)/n$ for $y(s)$ and $K_n(s)$ for $k(s)$.
A nonparametric plug-in option is to substitute 
$\d\hatt H(s)=\d N(s)/Y(s)$ for $h(s,\theta)\,\d s$ everywhere in (2.3),
and again using $Y(s)/n$ and $K_n(s)$. It is shown in Hjort (1990,
section 3) that each of these choices, or indeed combination
of these choices, lead to consistent estimators of $\kappa(t)^2$.  

The natural parametric estimator of $\Sigma$ is 
$$\hatt\Sigma_{\rm pm}=\int_0^\tau\psi(s,\hatt\theta)\psi(s,\hatt\theta)^\tr 
	\{Y(s)/n\}h(s,\hatt\theta)\,\d s
=n^{-1}\sumjn\int_0^{t_j}\psi(s,\hatt\theta)\psi(s,\hatt\theta)^\tr \,
	h(s,\hatt\theta)\,\d s, \eqno(2.4)$$
and is easy to calculate in cases where the integrals can be done
explicitly. The natural nonparametric alternative is 	
$$\eqalign{
\hatt\Sigma_{\rm np}
&=\int_0^\tau\psi(s,\hatt\theta)\psi(s,\hatt\theta)^\tr 
{Y(s)\over n}\,{\d N(s)\over Y(s)} \cr
&=n^{-1}\sum_{j\colon \delta_j=1}
	\psi(t_j,\hatt\theta)\psi(t_j,\hatt\theta_j)^\tr  
=n^{-1}\sumjn\psi(t_j,\hatt\theta)\psi(t_j,\hatt\theta)^\tr \delta_j,} $$
which is easier to calculate where the integrals do not allow
explicit expressions. 
These are both consistent estimates, under model conditions. 	

\subsection
{\csc 2.3. NLH-plots, Type A.}
Type A of (1.4) uses $K_n(s)=1$, and has 
$$\kappa_A(t)^2=\int_0^t{h(s,\theta)\over y(s)}\,\d s
-\Bigl(\int_0^th(s,\theta)\psi(s,\theta)\,\d s\Bigr)^\tr 
\Sigma^{-1}\Bigl(\int_0^th(s,\theta)\psi(s,\theta)\,\d s\Bigr). $$
The parametric plug-in option is 
$$\hatt\kappa_A(t)^2=\int_0^t{n\over Y(s)}h(s,\hatt\theta)\,\d s
-H^*(t,\hatt\theta)^\tr \hatt\Sigma_{\rm pm}^{-1}H^*(t,\hatt\theta), \eqno(2.5)$$
where we write $H^*(t,\theta)$ for ${\dell\over \dell\theta}H(t,\theta)
=\int_0^th(s,\theta)\psi(s,\theta)\,\d s$,
for notational convenience. 
The first integral is easiest to calculate 
by adding over the ordered $(t_{j-1},t_j]$ intervals to the left 
of time point $t$, since $Y(s)$ is constant over each of these. 
One then gets a sum of type 
$\sum_{j\colon t_j\le t}\{n/Y(t_{j-1})\}\{H(t_j,\hatt\theta)
\allowbreak-H(t_{j-1},
\hatt\theta)\}$ plus a similar term for the remaining $(t_l,t]$ interval,
where $t_l$ is the last $t_j$ before $t$. 
The nonparametric option is 
$$\hatt\kappa_A(t)^2=\int_0^t{n\,\d N(s)\over Y(s)^2}
	-\Bigl(\int_0^t\psi(s,\hatt\theta){\d N(s)\over Y(s)}\Bigr)^\tr 
	\hatt\Sigma_{\rm np}^{-1}
	\Bigl(\int_0^t\psi(s,\hatt\theta){\d N(s)\over Y(s)}\Bigr).$$
Note that the first term is $n$ times the usual variance estimator 
for the Nelson--Aalen estimator, see ABGK (chapter IV). 
The second integral is the finite sum 
$\sum_{t_j\le t}\psi(t_j,\hatt\theta)\delta_j/Y(t_j)$. 

\subsection
{\csc 2.4. NLH-plots, Type B.}
Next consider Type B of (1.4), which uses $K_n(s)=Y(s)/n$, and has 
$$\kappa_B(t)^2=\int_0^ty(s)h(s,\theta)\,\d s
-\Bigl(\int_0^ty(s)h(s,\theta)\psi(s,\theta)\,\d s\Bigr)^\tr 
\Sigma^{-1}\Bigl(\int_0^ty(s)h(s,\theta)\psi(s,\theta)\,\d s\Bigr). $$
The parametric choice becomes 
$$\hatt\kappa_B(t)^2=n^{-1}\sumjn H(t_j\wedge t,\hatt\theta)
-\Bigl(n^{-1}\sumjn H^*(t_j\wedge t,\hatt\theta)\Bigr)^\tr \hatt\Sigma_{\rm pm}^{-1}
 \Bigl(n^{-1}\sumjn H^*(t_j\wedge t,\hatt\theta)\Bigr), \eqno(2.6)$$
while the nonparametric version is 
$$\hatt\kappa_B(t)^2=N(t)/n
-\Bigl(n^{-1}\sum_{t_j\le t}\psi(t_j,\hatt\theta)\delta_j\Bigr)^\tr 
\hatt\Sigma_{\rm np}^{-1}
\Bigl(n^{-1}\sum_{t_j\le t}\psi(t_j,\hatt\theta)\delta_j\Bigr). $$

\subsection
{\csc 2.5. NLH-plots, Type C.}
Finally consider the third choice in (1.4), 
which employs a weight function of the $K_n(s)=\{Y(s)/n\}G_n(s)$.
Suppose for example that an omnibus goodness of fit test for 
the $h(s)=h(s,\theta)$ model is sought, that at the same time
has good detection power for neighbouring alternatives of the form
$h(s,\theta,\gamma)$, where $h(s,\theta)$ is the special
case $h(s,\theta,\gamma_0)$. Then an optimal choice for weight
function in (1.3) can be shown to be of the form 
$$K_n(s)=n^{-1}Y(s)\Big\{\phi(s,\hatt\theta)-\psi(s,\hatt\theta)^\tr 
	\hatt\Sigma_{\rm pm}^{-1}\int_0^\tau n^{-1}Y(u)\phi(u,\hatt\theta)
	\psi(u,\hatt\theta)h(u,\hatt\theta)\,\d u\Big\}, \eqno(2.7)$$
where $\phi(s,\theta)={\dell\over \dell\gamma}\log h(s,\theta,\gamma_0)$.
There is also a nonparametric alternative. 
See Hjort (1990, section 5) and Koning (1991, section 4) for details,
and 3.2 and 5.2 below for illustrations. 

The actual implementation of this case is reasonably similar to
that of Type B above. One uses
$${\rm NLH}_C(t)=n^{-1/2}\Bigl\{\int_0^tG_n(s)\,\d N(s)
-\sumjn\int_0^{t_j\wedge t}G_n(s)h(s,\hatt\theta)\,\d s\Bigr\}
/\hatt\kappa_C(t), $$
where 
$$\eqalign{
\hatt\kappa_C(t)^2
&=n^{-1}\sumjn\int_0^{t_j\wedge t}
G_n(s)^2h(s,\hatt\theta)\,\d s \cr
&\qquad -\Bigl(n^{-1}\sumjn\int_0^{t_j\wedge t}G_n(s)\psi(s,\hatt\theta)
	h(s,\hatt\theta)\,\d s\Bigr)^\tr 
\hatt\Sigma_{\rm pm}^{-1} \cr
& \qquad\qquad \Bigl(n^{-1}\sumjn\int_0^{t_j\wedge t}G_n(s)\psi(s,\hatt\theta)
	h(s,\hatt\theta)\,\d s\Bigr).  \cr}$$
There is also a nonparametric alternative for $\kappa(t)$ estimation,
for cases where calculating the above becomes too complicated. 	
	
\subsection
{\csc 2.6. Discussion.}
The parametric plug-in versions are more statistically precise
estimators than the nonparametric ones, and are chosen whenever they 
are not too cumbersome to compute.
This means choosing Type A plots with (2.5) and Type B plots with (2.6). 
For some models, like the gamma and the log-normal, 
these can only be computed with numerical integration, however, 
in which case the nonparametric options are easier, 
involving only finite sums of explicit terms. 
We also note that positivity of the various $\kappa(.)$-estimate functions
as given here is guaranteed, unlike for other choices that may seem
natural and are as permissible from the asymptotic statistics point of view.
See Remark 8E. 

\bigskip
{\bf 3. Special models.}

\subsection
{\csc 3.1. A completely specified hazard function.} 
Sometimes one wants to compare the life-time distribution for 
a group of individuals with some established norm, 
say one with hazard rate $h_0(.)$. In this case 
the limit process of $D_n(.)$ is simpler than in the case with estimated
parameters. It is $D(t)=\int_0^t\{k(s)/y(s)\}\,\d V(s)$,
which is simply time-transformed Brownian motion, 
$W(\kappa(t)^2)$, where $\kappa(t)^2=\int_0^t\{k(s)^2/y(s)\}h_0(s)\,\d s$.
The test curve becomes
$${\rm NLH}(t)=\sqrt{n}\int_0^tK_n(s)\{\d\hatt H(s)-h_0(s)\,\d s\}
	\Big/\Bigl\{\int_0^tK_n(s)^2\{n/Y(s)\}
	h_0(s)\,\d s\Bigr\}^{1/2}, \eqno(3.1)$$
with appropriate simplifications for Type A and Type B. 
Its limit is a time-transformed normalised Brownian motion, see Remark 8A. 

\subsection
{\csc 3.2. Testing the constant hazard rate model.}
Let the model be a constant rate $h(s,\theta)=\theta$.
Then $H(t,\theta)=\theta t$ and $\psi(s,\theta)=1/\theta$.
The ML estimator solves $\int_0^\tau\{\d N(s)-Y(s)\theta\,\d s\}=0$,
that is, 
$$\hatt\theta={N(\tau)\over \int_0^\tau Y(s)\,\d s}
	={\sumjn\delta_j\over \sumjn t_j}. $$
In this case both the parametric and nonparametric
estimates of
$$ \Sigma=\sigma^2=\int_0^\tau y(s)\{1/\theta^2\}\theta\,\d s $$
become equal to $\hatt\sigma^2=n^{-1}(\sumjn t_j)^2/\sumjn\delta_j$.
Note also that $\hatt\sigma^2=1/\hatt\theta^2$ in the case of 
non-censored data.  	

The NLH-plot of Type A for exponentiality is 
$$\sqrt{n}\{\hatt H(t)-\hatt\theta t\}/\hatt\kappa_A(t), \eqno(3.2)$$
where the parametric plug-in estimator for the variance is 
$$\hatt\kappa_A(t)^2=\int_0^t{n\over Y(s)}\hatt\theta\,\d s-t^2/\hatt\sigma^2
=\hatt\theta\Bigl\{\int_0^t{n\over Y(s)}\,\d s
	-{n\over N(\tau)}\hatt\theta t^2\Bigr\}. $$
The nonparametric alternative is 
$$\hatt\kappa_A(t)^2=\int_0^t{n\,\d N(s)\over Y(s)^2}-\Bigl\{\int_0^t
{1\over \hatt\theta}{\d N(s)\over Y(s)}\Bigr\}^2/\hatt\sigma^2
=\int_0^t{n\,\d N(s)\over Y(s)^2}-\{\hatt H(t)/\hatt\theta\}^2/\hatt\sigma^2.$$	
Similarly there is a NLH-plot of Type B for exponentiality:
$$n^{-1/2}\Bigl\{N(t)-\int_0^t Y(s)\hatt\theta\,\d s\Bigr\}/\hatt\kappa_B(t)
	=n^{-1/2}\Bigl\{N(t)-\sumjn \hatt\theta(t_j\wedge t)\Bigr\}
	/\hatt\kappa_B(t). \eqno(3.3)$$
The parametric choice for variance estimation is 
$$\hatt\kappa_B(t)^2=n^{-1}\sumjn(t_j\wedge t)\hatt\theta-
\Bigl(n^{-1}\sumjn(t_j\wedge t)\Bigr)^2/\hatt\sigma^2, $$
while the nonparametric version is
$n^{-1}N(t)-(n^{-1}N(t)/\hatt\theta)^2/\hatt\sigma^2$. 

Let us finally include a version of the Type C plots.
If an overall test for exponentiality is sought that at the
same time is good at detecting departures in the direction of
Weibullness, then the general recipe described in 2.5 above leads to 
the tailor-made weight function $K_n(s)=\{Y(s)/n\}G_n(s)$, with
$$G_n(s)=\log s-\hatt\phi=\log s-{\int_0^\tau Y(s)\log s\,\d s
\over \int_0^\tau Y(s)\,\d s}=\log s-{\sumjn(t_j\log t_j-t_j)
\over \sumjn t_j}.$$
The result is 
$${\rm NLH}_C(t)=n^{-1/2}\Bigl\{\int_0^t (\log s-\hatt\phi)\,\d N(s)
-\sumjn\int_0^{t_j\wedge t}(\log s-\hatt\phi)\hatt\theta\,\d s\Bigr\}
	/\hatt\kappa_C(t),$$
where 
$$\hatt\kappa_C(t)^2=n^{-1}\sumjn\int_0^{t_j\wedge t}(\log s-\hatt\phi)^2
	\hatt\theta\,\d s
-\Bigl\{n^{-1}\sumjn\int_0^{t_j\wedge t}(\log s-\hatt\phi)\,\d s\Bigr\}^2
	/\hatt\sigma^2.$$
And of course the integrals here can be explicitly evaluated. 

These test curves are illustrated in examples 9.1--9.3. 

Note that the case of a model $h(s,\theta)=\theta h_0(s)$ 
specifying proportionality to a specified $h_0(s)$ can be
treated in the same way, with small adjustments. 

\subsection 
{\csc 3.3. A class of two-parameter models.} 
Suppose $h(s,\theta,\beta)=\theta h_0(s,\beta)$ for a specified
$h_0$ function in terms of a single extra $\beta$ parameter. 
Then
$$ H(t,\theta,\beta)=\theta\int_0^th_0(s,\beta)\,\d s=\theta H_0(t,\beta), $$
and $\log h$ has derivatives 
$\psi(s,\theta,\beta)=(\theta^{-1},\psi_0(s,\beta))$. 
The ML estimates maximise 
$$L(\theta,\beta)=\int_0^\tau\bigl[\{\log\theta
+\log h_0(s,\beta)\}\,\d N(s)
	-Y(s)\theta h_0(s,\beta)\,\d s\bigr]. \eqno(3.4)$$
We see that 
$\hatt\theta(\beta)=N(\tau)/\int_0^\tau Y(s)h_0(s,\beta)\,\d s$,
and $\hatt\beta$ maximises the resulting profile log-likelihood,
so this can be made into a one-dimensional problem. See the computational
notes of the Appendix. 
Next we need to estimate the matrix
$$\Sigma=\int_0^\tau\mtrix{1/\theta \cr \psi_0(s,\beta) \cr}
\mtrix{1/\theta \cr \psi_0(s,\beta) \cr}^\tr y(s)\theta h_0(s,\beta)\,\d s.$$
The parametric estimate is 
$$\hatt\Sigma_{\rm pm}=n^{-1}\sumjn\int_0^{t_j}
\mtrix{1/\hatt\theta^2 &\psi_0(s,\hatt\beta)/\hatt\theta \cr
	\psi_0(s,\hatt\beta)/\hatt\theta & \psi_0(s,\hatt\theta)^2 \cr}
	\hatt\theta h_0(s,\hatt\beta)\,\d s, \eqno(3.5)$$
and in many cases of interest the integrals can be evaluated explicitly. 
The nonparametric version is the finite sum 
$$\hatt\Sigma_{\rm np}=n^{-1}\sumjn 
\mtrix{1/\hatt\theta^2 & \psi_0(t_j,\hatt\beta)/\hatt\theta \cr	
\psi_0(t_j,\hatt\beta)/\hatt\theta & \psi_0(t_j,\hatt\beta)^2 \cr}\delta_j.$$

NLH-plots of Type A plots are now of the form $\sqrt{n}\{\hatt H(t)
-\hatt\theta H_0(t,\hatt\beta)\}/\hatt\kappa_A(t)$ with 
$$\eqalign{
\hatt\kappa_A(t)^2
=\int_0^t{n\over Y(s)}\hatt\theta h_0(s,\hatt\beta)\,\d s 
-\mtrix{H_0(t,\hatt\beta) \cr \hatt\theta H_0^*(t,\hatt\beta) \cr}^\tr 
\hatt\Sigma_{\rm pm}^{-1}
\mtrix{H_0(t,\hatt\beta) \cr \hatt\theta H_0^*(t,\hatt\beta) \cr},} \eqno(3.6)$$
in which $H_0^*(t,\beta)={\dell\over \dell\beta}H_0(t,\beta)$. 
Similarly NLH-plots of Type B are of the form
$$n^{-1/2}\{N(t)-\int_0^tY(s)\hatt\theta 
   h_0(s,\hatt\beta)\,\d s\}/\hatt\kappa_B(t), $$
with 
$$\eqalign{
\hatt\kappa_B(t)^2
&=n^{-1}\sumjn\hatt\theta H_0(t_j\wedge t,\hatt\beta) \cr
&\qquad -\mtrix{n^{-1}\sumjn H_0(t_j\wedge t,\hatt\beta) \cr
n^{-1}\sumjn \hatt\theta H_0^*(t_j\wedge t,\hatt\beta) \cr}^\tr 
\hatt\Sigma_{\rm pm}^{-1}
\mtrix{n^{-1}\sumjn H_0(t_j\wedge t,\hatt\beta) \cr
n^{-1}\sumjn \hatt\theta H_0^*(t_j\wedge t,\hatt\beta) \cr}.} \eqno(3.7)$$
The nonparametric options for variance estimation 
are also available and sometimes simpler to compute.

\subsection
{\csc 3.4. Weibull.} 
Here $h(s,\theta,\beta)=\theta\beta s^{\beta-1}$ with cumulative 
hazard $H(t,\theta,\beta)=\theta t^\beta$. 
The ML estimates maximise
$\sumjn[\{\log\theta+\log\beta+(\beta-1)\log t_j\}\delta_j
	-\theta t_j^{\beta}]$. 
The construction above leads after some calculations to 
$$\hatt\Sigma_{\rm pm}=\mtrix{n^{-1}\hatt\theta^{-1}\sumjn t_j^{\hatt\beta} &
n^{-1}\hatt\beta^{-1}\sumjn t_j^{\hatt\beta}\log t_j^{\hatt\beta} \cr  
n^{-1}\hatt\beta^{-1}\sumjn t_j^{\hatt\beta}\log t_j^{\hatt\beta} 
& n^{-1}\hatt\theta\hatt\beta^{-2} 
\sumjn t_j^{\hatt\beta}\{1+(\log t_j^{\hatt\beta})^2\} \cr}.$$
We are now in a position to properly define NLH-plots for Weibullness.
{\sl Type A} uses 
$${\rm NLH}_A(t)=\sqrt{n}\{\hatt H(t)
-\hatt\theta t^{\hatt\beta}\}/\hatt\kappa_A(t), \eqno(3.8)$$ 
where 
$$\hatt\kappa_A(t)^2=\int_0^t{n\over Y(s)}\hatt\theta 
\hatt\beta s^{\hatt\beta-1}\,\d s
-\mtrix{t^{\hatt\beta} \cr \hatt\theta\hatt\beta^{-1} 
t^{\hatt\beta}\log t^{\hatt\beta} \cr}^\tr 
\hatt\Sigma_{\rm pm}^{-1}
\mtrix{t^{\hatt\beta} \cr \hatt\theta\hatt\beta^{-1} 
t^{\hatt\beta}\log t^{\hatt\beta} \cr}. $$
And the NLH-plot of {\sl Type B} is
$$n^{-1/2}\Bigl\{N(t)-\int_0^tY(s)\hatt\theta 
\hatt\beta s^{\hatt\beta-1}\,\d s\Bigr\}/\hatt\kappa_B(t)
=n^{-1/2}\Bigl\{N(t)-\sumjn\hatt\theta (t_j\wedge t)^{\hatt\beta}\Bigr\}/
\hatt\kappa_B(t), \eqno(3.9)$$
where the denominator is the square root of 
$$\eqalign{
&n^{-1}\sumjn\hatt\theta(t_j\wedge t)^{\hatt\beta} \cr
&\quad -\mtrix{n^{-1}\sumjn(t_j\wedge t)^{\hatt\beta} \cr
n^{-1}\sumjn\hatt\theta(t_j\wedge t)^{\hatt\beta}\log(t_j\wedge t)}^\tr 
\hatt\Sigma_{\rm pm}^{-1}
\mtrix{n^{-1}\sumjn(t_j\wedge t)^{\hatt\beta} \cr
n^{-1}\sumjn\hatt\theta(t_j\wedge t)^{\hatt\beta}\log(t_j\wedge t)}.}$$
There are also nonparametric alternatives to these variance estimators. 


An illustration is given in example 9.2. 
\eject 

\subsection
{\csc 3.5. Frailty models.} 
Suppose each individual has his own constant hazard rate,
and that these are distributed in the population under study 
as a gamma distribution with mean $\theta$ and variance $\theta\beta$. 
Then the hazard rate for observed individuals is of the form 
$h(s,\theta,\beta)=\theta/(1+\beta s)$, nonincreasing a priori. 
The cumulative increases logarithmically. 
We call this the simple frailty model. 

The ML estimates are found by maximising 
$\sumjn[\{\log\theta-\log(1+\beta t_j)\}\delta_j
	-\theta\beta^{-1}\log(1+\beta t_j)]$.
With some efforts the entries of $\hatt\Sigma_{\rm pm}$ are found:
$$\eqalign{
\hatt\Sigma_{1,1}
&=n^{-1}\sumjn\hatt\theta^{-1}\hatt\beta^{-1}\log(1+\hatt\beta t_j), \cr 
\hatt\Sigma_{1,2}
&=-n^{-1}\sumjn\hatt\beta^{-2}\{\log(1+\hatt\beta t_j)
	+(1+\hatt\beta t_j)^{-1}-1\}, \cr
\hatt\Sigma_{2,2}
&=n^{-1}\sumjn\hatt\theta\hatt\beta^{-3}\{\log(1+\hatt\beta t_j)
	+2(1+\hatt\beta t_j)^{-1}-\half(1+\hatt\beta t_j)^{-2}
	-\hbox{$3\over2$}\}. \cr}$$

NLH-plots to check the fit of the model can now be properly defined,
parallelling (3.8) and (3.9). The Type A plot is 
$${\rm NLH}_A(t)=\sqrt{n}\{\hatt H(t)
-\hatt\theta\hatt\beta^{-1}\log(1+\hatt\beta t)\}
	/\hatt\kappa_A(t), \eqno(3.10)$$
where 
$$\hatt\kappa_A(t)^2=\int_0^t{n\over Y(s)}{\hatt\theta\over 1+\hatt\beta s}
\,d s-\hatt C(t)^\tr \hatt\Sigma_{\rm pm}^{-1}\hatt C(t), $$	
in which 
$$\hatt C(t)=\mtrix{
\hatt\beta^{-1}\log(1+\hatt\beta t) \cr
\hatt\theta\hatt\beta^{-2}\{{\hatt\beta t\over 1+\hatt\beta t}
	-\log(1+\hatt\beta t)\} \cr}. $$
And the Type B plot uses 
$$n^{-1/2}\Bigl\{N(t)-\int_0^tY(s){\hatt\theta\over 1+\hatt\beta s}\,\d s\Bigr\}
/\hatt\kappa_B(t)
=n^{-1/2}\Bigl\{N(t)-\sumjn\hatt\theta\hatt\beta^{-1}
\log(1+\hatt\beta(t_j\wedge t))\Bigr\}/\hatt\kappa_B(t), \eqno(3.11)$$
where 
$$\hatt\kappa_B(t)^2=n^{-1}\sumjn\hatt\theta\hatt\beta^{-1}
\log(1+\hatt\beta(t_j\wedge t))
-\Bigl(n^{-1}\sumjn\hatt C(t_j\wedge t)\Bigr)^\tr \hatt\Sigma_{\rm pm}^{-1}
\Bigl(n^{-1}\sumjn\hatt C(t_j\wedge t)\Bigr). $$

An illustration is given in example 9.3.

This simple frailty model should by its motivation and construction have
$\beta\ge0$. If the likelihood is maximised only in this region 
then the ML estimator will be equal to zero with positive probability,
and one encounters the more involved `corner asymptotics' problems
associated with zero not being an inner point of the parameter space.
In particular the plots would not be approximately standard normal
any more. We avoid these difficulties by allowing an expanded
parameter space $\beta>-\eps$, where $1/\eps$ is twice the largest
observed $t_j$. And of course if the Nelson--Aalen plot for data
suggests an increasing hazard rate one should immediately abandon
the frailty model. 

There are more general models based on frailty distributions and that
have found uses in survival analysis and demography, see Aalen (1992). 
One important class of models is as follows. 
Let $\lambda(t)$ be a hazard rate function with cumulative 
$\Lambda(t)=\int_0^t\lambda(s)\,\d s$. The individuals in the population
under study have all hazard rates of the type $Z\lambda(t)$,
but with $Z$, the unobservable frailty factor, varying from individual
to individual. When $Z$ has the particular compound Poisson 
distribution considered by Aalen, with certain parameters $\alpha$ and 
$\delta$, then the life-time of a randomly chosen individual has 
$$\eqalign{{\rm hazard\ rate\ }h(t)
&={\lambda(t)\over \{1+(\delta/\alpha)\Lambda(t)\}^\alpha}, \cr
{\rm with\ cumulative\ }H(t)&={\alpha\over \alpha-1}{1\over \delta}
        \Bigl[1-\Bigl(1+{\delta\over \alpha}
                \Lambda(t)\Bigr)^{1-\alpha}\Bigr]. \cr} \eqno(3.12)$$
The last formula is valid for $\alpha\not=1$. For $\alpha=1$
one has $H(t)=\delta^{-1}\log\{1+\delta\Lambda(t)\}$,
and the special case above is of this form.            
If the underlying $\lambda(.)$ is parametrised with two parameters
then the scheme above leads to a four-parameter hazard model, for example. 
And the validity of these can all be tested with the NLH plot machinery.
See example 9.6 for an illustration.

\subsection 
{\csc 3.6. Gompertz.} 
Here $h(s)=\theta\exp(\beta s)$ on $[0,\tau]$. The ML estimators maximise
$\sumjn[(\log\theta+\beta t_j)\delta_j
	-\theta\beta^{-1}\{\exp(\beta t_j)-1\}]$, 
and this function is actually concave in $(\log\theta,\beta)$.	
The entries of $\hatt\Sigma_{\rm pm}$ are
$$\eqalign{
\hatt\Sigma_{1,1}&=n^{-1}\hatt\theta^{-1}\sumjn
	\hatt\beta^{-1}\{\exp(\hatt\beta t_j)-1\}, \cr
\hatt\Sigma_{1,2}&=n^{-1}\sumjn
	\hatt\beta^{-2}\{(\hatt\beta t_j-1)\exp(\hatt\beta t_j)+1\}, \cr
\hatt\Sigma_{2,2}&=n^{-1}\hatt\theta\sumjn
\hatt\beta^{-3}[\{(\hatt\beta t_j)^2-2\hatt\beta t_j+2\}
	\exp(\hatt\beta t_j)-2]. \cr}$$	
And quite similarly to the cases above the A-plot and the B-plot take the forms
$$\sqrt{n}[\hatt H(t)-\hatt\theta\hatt\beta^{-1}
	\{\exp(\hatt\beta t)-1\}]/\hatt\kappa_A(t) 
{\rm\quad and\quad}	
n^{-1/2}\Bigl\{N(t)-\int_0^tY(s)\hatt\theta\exp(\hatt\beta s)\,\d s\Bigr\}
	/\hatt\kappa_B(t), $$
in which 
$$\hatt\kappa_A(t)^2=\int_0^t{n\over Y(s)}\hatt\theta\exp(\hatt\beta s)\,\d s
	-\hatt C(t)^\tr \hatt\Sigma_{\rm pm}^{-1}\hatt C(t) $$
and
$$\hatt\kappa_B(t)^2=n^{-1}\sumjn\hatt\theta\hatt\beta^{-1}
\{\exp(\hatt\beta(t_j\wedge t))-1\}	
-\Bigl(n^{-1}\sumjn\hatt C(t_j\wedge t)\Bigr)^\tr \hatt\Sigma_{\rm pm}^{-1}
\Bigl(n^{-1}\sumjn\hatt C(t_j\wedge t)\Bigr). $$
This time
$$\hatt C(t)=\mtrix{
\hatt\beta^{-1}\{\exp(\hatt\beta t)-1\} \cr
\hatt\theta\hatt\beta^{-2}
	\{(\hatt\beta t-1)\exp(\hatt\beta t)+1\} \cr}. $$

\subsection
{\csc 3.7. Gamma.}
The gamma distribution has density 
$$ f(t,\alpha,\theta)=\{\theta^\alpha/\Gamma(\alpha)\}
	t^{\alpha-1}\exp(-\theta t), $$
and is a useful class to work with in connection with life history data. 
In some situations a priori reasons could suggest a known integer value 
for the shape parameter $\alpha$, in which case one also uses the name 
Erlang distribution. The constant hazard model corresponds to $\alpha=1$,
and with $\alpha=3$, for example, the hazard function 
$h(t,\alpha,\theta)=f(t,\alpha,\theta)/\{1-F(t,\alpha,\theta)\}$ becomes
$\theta\,\half\theta^2t^2/\{1+\theta t+\half\theta^2t^2\}$, 
smoothly increasing from 0 to its asymptote $\theta$. 
It is not very difficult to construct NLH-plots for testing the
fit of this one-parameter hazard model, or the other models 
with known integer $\alpha$. 

The general two-parameter model also fits into our general framework, 
but its analysis is more cumbersome than for the
previous examples in that its hazard function and its derivatives 
w.r.t.~model parameters involve special mathematical
functions like the incomplete gamma integral. 
The cumulative distribution is 
$F(t,\alpha,\theta)=F_0(\theta t,\alpha)$, where 
$F_0(t,\alpha)=\int_0^t\Gamma(\alpha)^{-1}u^{\alpha-1}e^{-u}\,\d u$,
for example. This function is available in \Splus{} and other
computer package libraries. Let us 
present the necessary formulae. First, the ML estimators maximise 
$$\sum_{j\colon\delta_j=1}
\{\alpha\log(\theta t_j)-\log\Gamma(\alpha)-\theta t_j\}
+\sum_{j\colon\delta_j=0}\log\{1-F_0(\theta t_j,\alpha)\}. $$
In this case the parametric options for estimating $\Sigma$ and 
the $\kappa(.)$ functions are complicated, although they can be managed
via numerical integrations, and the nonparametric options are easier.
The two plots to test gamma-ness become
$$\sqrt{n}[\hatt H(t)+\log\{1-F_0(\hatt\theta t,\hatt\alpha)\}]/\hatt\kappa_A(t)
{\rm\ and\ }
n^{-1/2}\Bigl[N(t)+\sumjn\log\{1-F_0(\hatt\theta(t_j\wedge t),
	\hatt\alpha)\}\Bigr]/\hatt\kappa_B(t), $$
where the denominators involve $\hatt\Sigma_{\rm np}$ 
and are the nonparametric options as given generally in 2.3 and 2.4.
These are finite sums involving the model's $\psi(t,\alpha,\theta)$ 
function, with components 
$$\eqalign{\psi_\alpha(t,\alpha,\theta)
&=\log(\theta t)-\psi(\alpha)+\{1-F_0(\theta t,\alpha)\}^{-1}
\int_0^t\{\log(\theta s)-\psi(\alpha)\}f(s,\alpha,\theta)\,\d s \cr
&=\log(\theta t)-\psi(\alpha)+\{1-F_0(\theta t,\alpha)\}^{-1}
\{F_0^*(\theta t,\alpha)-\psi(\alpha)F_0(\theta t,\alpha)\}, \cr
\psi_\theta(t,\alpha,\theta)
&=\theta^{-1}(\alpha-\theta t)+\{1-F_0(\theta t,\alpha)\}^{-1}
\int_0^t\theta^{-1}(\alpha-\theta s)f(s,\alpha,\theta)\,\d s \cr
&=\theta^{-1}(\alpha-\theta t)+\{1-F_0(\theta t,\alpha)\}^{-1}
\theta^{-1}(\theta t)^\alpha\exp(-\theta t)/\Gamma(\alpha). \cr}$$
Here $\psi(\alpha)=\Gamma^\tr (\alpha)/\Gamma(\alpha)$, and 
in addition to $F_0(t,\alpha)$ one needs its relative 
$F_0^*(t,\alpha)=\int_0^t\Gamma(\alpha)^{-1}\log u\allowbreak 
u^{\alpha-1}e^{-u}\,\d u$. 

\subsection
{\csc 3.8. Other models.} 
Constructing NLH curves for other parametric models for survival data
should not be difficult given the general machinery and the examples
above. Some useful models include the Gompertz--Makeham one with
$h(s)=a+b\exp(c s)$, the lognormal, the log-logistic, 
the inverse normal, and the three-parameter 
$h(s)=\exp(\beta_0+\beta_1s+\beta_2s^2)$, 
capable of representing a wide variety of hazard curves on $[0,\tau]$. 

\bigskip
{\bf 4. NLH-plots for the parametric Cox regression model.}
Suppose covariate information in the form of features or 
measurements $z_j=(z_{j,1},\ldots,z_{j,p})^\tr $ are available for 
individual $j$, in addition to the possibly right-censored life-time
information $(t_j,\delta_j)$. The widely used semiparametric Cox
regression model postulates that individual $j$ has hazard rate of the form 
$h_j(s)=h_0(s)\exp(\beta^\tr z_j)$, but without further specifying $h_0(.)$,
the hazard rate for individuals with covariate zero. 
NLH-plots can be constructed to check the validity of 
a {\it parametric} Cox model, where 
$$h_j(s)=h(s,\theta)\exp(\beta_1z_{j,1}+\cdots+\beta_pz_{j,p})
	\quad {\rm for\ }j=1,\ldots,n. \eqno(4.1)$$
If such a parametric form for the baseline hazard can be validated
it makes for a better understanding of the survival mechanisms
under study, and also makes it possible to estimate the
$\beta$ parameters with increased precision. 

We shall be content to illustrate the use of our plots 
for the important special case where the baseline hazard is 
a constant $\theta$. This is sometimes a quite effective model. 
We also assume that the covariates are constant in time.
Generalisations to time-dependent covariates and to other
$h(s,\theta)$ models are reasonably straightforward, 
in view of the theory developed in Hjort (1990, section 6). 

The ML estimates are computed from the log-likelihood 
$$\sumjn\int_0^\tau\{(\log\theta+\beta^\tr z_j)\,\d N_j(s)
-Y_j(s)\theta\exp(\beta^\tr z_j)\,\d s\}
=\sumjn\{(\log\theta+\beta^\tr z_j)\delta_j-\theta\exp(\beta^\tr z_j)\,t_j\}.$$	
Here $Y_j(s)=I\{t_j\ge s\}$ is the at risk indicator 
and $N_j(t)=I\{t_j\le t,\delta_j=1\}$ the 0--1 counting process for no.~$j$,
summing to the $N=\sumjn N_j$ counting process. 
Define 
$$\eqalign{
\hatt\Sigma_{1,1}
   &=n^{-1}\hatt\theta^{-1}\sumjn\exp(\hatt\beta^\tr z_j)\,t_j,\, \cr
\hatt\Sigma_{2,1}
   &=n^{-1}\sumjn z_j\exp(\hatt\beta^\tr z_j)\,t_j,\, \cr
\hatt\Sigma_{2,2}
   &=n^{-1}\hatt\theta\sumjn z_jz_j^\tr \exp(\hatt\beta^\tr z_j)\,t_j.}$$
These are the blocks of a consistent estimator $\hatt\Sigma$ for
the inverse variance matrix for the limit distribution of 
$\sqrt{n}(\hatt\theta-\theta,\hatt\beta-\beta)$, under the (4.1) model,
as can be shown from more general results of Hjort (1990, section 6). 

A NLH-plot of Type A takes as its starting point
$$D_n(t)=\sqrt{n}\{\hatt H(t)-\hatt\theta t\}
=\sqrt{n}\Bigl\{\int_0^t{\d N(s)\over \sumjn Y_j(s)\exp(\hatt\beta^\tr z_j)}
	-\hatt\theta t\Bigr\}. \eqno(4.2)$$
One might also argue in favour of plugging in the maximum partial likelihood
estimator $\tilda\beta$ here, but the following formulae relate 
to the ML estimate from the parametric regression. 
Several natural consistent estimates for its limit variance can be constructed
along the general lines of Hjort (1990, section 6). One version is
as follows. Let $R(s,\beta)=n^{-1}\sumjn Y_j(s)\exp(\beta^\tr z_j)$, 
$R_{(1)}(s,\beta)=n^{-1}\sumjn Y_j(s)z_j\exp(\beta^\tr z_j)$,
and finally $E(s,\beta)=R_{(1)}(s,\beta)/R(s,\beta)$. These are 
next used to define 
$$\hatt C_A(t)=\mtrix{t\cr \int_0^t E(s,\hatt\beta)\hatt\theta\,\d s \cr},
\quad \hatt C_B(t)=\mtrix{\int_0^tR(s,\hatt\beta)\,\d s \cr
\int_0^tR_{(1)}(s,\hatt\beta)\,\hatt\theta\,\d s \cr}.$$ 
In the end define 
$$\hatt\kappa_A(t)^2=\int_0^t{\hatt\theta\,\d s\over R(s,\hatt\beta)}
-\hatt C_A(t)^\tr \hatt\Sigma^{-1}\hatt C_A(t), $$
and use ${\rm NLH}_A(t)=\sqrt{n}\{\hatt H(t)-\hatt\theta t\}/\hatt\kappa_A(t)$.

Similarly a NLH-plot of Type B would use 
$$D_n(t)=n^{-1/2}\Bigl\{N(t)-\sumjn\int_0^tY_j(s)\hatt\theta
\exp(\hatt\beta^\tr z_j)\,\d s\Bigr\}
=n^{-1/2}\Bigl\{N(t)-\sumjn\hatt\theta\exp(\hatt\beta^\tr z_j)
	(t_j\wedge t)\Bigr\}.$$
In this case one may use 
$\hatt\kappa_B(t)^2=\int_0^tR(s,\hatt\beta)\hatt\theta\,\d s
-\hatt C_B(t)^\tr \hatt\Sigma^{-1}\hatt C_B(t)$, 
with $\hatt C_B(t)$ as above. 
The simplest way to compute the various integrals appearing here
is by noting that the integrands are constant over the ordered 
intervals $(t_{j-1},t_j]$, turning the integrals into finite sums.

We also note that the log-likelihood is concave 
once reparametrised with $\theta=\exp(\beta_0)$,
making computation relatively easy, and, incidentally,
making rigorous proofs of the necessary large-sample distribution
statements easier. 
That the required ${\rm NLH}_A(t)\arr_d\normal\{0,1\}$
and ${\rm NLH}_B(t)\arr_d\normal\{0,1\}$ hold, under 
standard regularity conditions, follows indeed from the general
results of Hjort (1990, section 6), but can be proven relatively
easily under almost minimal regularity assumptions when one uses methods of 
Hjort and Pollard (1993, section 7). 

See example 9.5 for an illustration of these techniques. 

\bigskip
{\bf 5. Detection power.}
Some analysis makes it possible to predict the behaviour of
NLH-plots outside model conditions. Information in this section
should also help in pinpointing exactly which aspects of a proposed 
model are wrong, in cases where the test curve determinedly wanders 
outside the $\pm1.96$ band. 

\subsection
{\csc 5.1. A fixed alternative.}
Suppose the true hazard is $h(.)$ rather than belonging to
the parametric class $h(.,\theta)$. Then one can establish 
$$D_n(t)/\sqrt{n}=\int_0^tK_n(s)\{\d\hatt H(s)-h(s,\hatt\theta)\,\d s\}
\arr_p\pi(t)=\int_0^tk(s)\{h(s)-h(s,\theta_0)\}\,d s, \eqno(5.1)$$
and there is uniform convergence in probability under suitable
assumptions, see Hjort (1990, section 5). For Type A plots this
is simply $H(t)-H(t,\theta_0)$. The $\theta_0$ parameter
value here is not `true' but rather `least false' in the sense 
of making $h(s,\theta_0)$ the best parametric approximant to $h(s)$,
see Hjort (1992) for details necessary to prove this and for 
the precise distance measure which is being minimised. 

This result shows that one can expect the NLH plot 
to decrease in regions where $h(t)<h(t,\theta_0)$ 
and increase in regions where $h(t)>h(t,\theta_0)$
(this assumes $k(t)$ positive, as with Type A and Type B). 
It also indicates that ${\rm NLH}(t)/\sqrt{n}$ is estimating
a well-defined discrepancy function $\pi(t)/\kappa_0(t)$,
and which is zero only if the model being tested is correct. 
Yet another pleasing theoretical consequence is that every
departure from the model will be detected by the NLH plots 
with probability 1 as $n$ grows; a non-zero $\pi(t)$ will 
send $|{\rm NLH}(t)|$ towards infinity with speed $\sqrt{n}|\pi(t)|$. 

If the model being tested is that of a constant hazard $\theta$,
then
$$\theta_0=\int_0^\tau y(s)h(s)\,\d s/\int_0^\tau y(s)\,\d s. $$
Each of the NLH plots of 3.2 will then tend to decrease 
in regions where the true hazard is less than $\theta_0$
and tend to increase when it is greater than $\theta_0$. 
If the true state of affairs is a Weibull with $\beta>1$, for example,
then the expected plot decreases up to certain $t_0$ and 
increases afterwards. The opposite happens if the true hazard
is a Weibull with $\beta<1$. 

\subsection
{\csc 5.2. Local alternatives.} 
Suppose $h_n(s)\doteq h(s,\theta)\{1+\phi(s,\theta)\delta/\sqrt{n}\}$. 
If one for example considers model alternatives of the form 
$h(s,\theta,\beta)$, where $\beta=\beta_0$ gives back $h(s,\theta)$, 
then $\phi(s,\theta)={\dell\over \dell\beta}\log h(s,\theta,\beta_0)$. 
Methods of Hjort (1990, section 5) can be used to prove
$${\rm NLH}(t)\arr_dD(t)/\kappa(t)+\delta a(t)/\kappa(t)
	\sim\normal\{\delta a(t)/\kappa(t),1\}, \eqno(5.2)$$
where 
$$\eqalign{
a(t)
&=\int_0^tk(s)h(s,\theta)\phi(s,\theta)\,\d s \cr 
& \quad -\Bigl(\int_0^tk(s)h(s,\theta)\psi(s,\theta)\,\d s\Bigr)^\tr \Sigma^{-1}
\Bigl(\int_0^\tau y(s)\phi(s,\theta)h(s,\theta)\psi(s,\theta)\,\d s\Bigr). }$$
This indicates the local detection power of the NLH plots. 
The optimal choice for weight function is as in (2.7),
leading to a test curve of Type C. 

For illustration take the constant $\theta$ model again,
and suppose the true hazard is $\theta\{1+\phi(s)\delta/\sqrt{n}\}$
for a suitable $\phi$ function. Then calculations show 
that ${\rm NLH}(t)$ is approximately a normal with variance one
and mean parameter 
$$\delta{\int_0^tk(s)\{\phi(s)-\bar\phi\}\,\d s\over
\bigl[\int_0^t\{k(s)^2/y(s)\}\,\d s
	-\{\int_0^tk(s)\,\d s\}^2/\int_0^\tau y(s)\,\d s\bigr]^{1/2}}, $$
in which $\bar\phi=\int_0^\tau y(s)\phi(s)\,\d s/\int_0^\tau y(s)\,\d s$. 
The optimal choice corresponds to $k(s)=y(s)\{\phi(s)-\bar\phi\}$.

For another illustration, consider the heterogeneity situation where
there is a `hidden covariate'. If the true hazard for individual $i$
is of the proportional hazard form $h_i(s)=h(s|z_i)=h_0(s)\exp(\beta z_i)$,
and the frailty factor $\exp(\beta z_i)$ has a gamma distribution with
mean 1 and variance $\sigma^2$ (which amounts to gamma parameters 
equal to $1/\sigma^2$ and $1/\sigma^2$), then calculations show that 
the hazard rate for a randomly selected individual is of the form
$h_0(s)/\{1+\sigma^2H_0(s)\}$, where $H_0(.)$ is the cumulative hazard
for $h_0(.)$. Suppose in particular that $h_i(s)=\theta\exp(\beta z_i)$
for an unobserved covariate of this form, with population variance 
$\sigma^2=\delta/\sqrt{n}$. Then the framework above is appropriate with
$h_n(s)\doteq\theta(1-\theta s\delta/\sqrt{n})$. If we further assume
that there is no censoring and the observation window is the full 
{half}line then calculations with $\phi(s,\theta)=-\theta s$, 
for the B plot, give 
$$a(t)=\theta t\exp(-\theta t) \quad {\rm and} \quad
	\kappa(t)^2=\{1-\exp(-\theta t)\}\exp(-\theta t). $$
In the end this leads to 	
$${\rm NLH}_B(t)\approx\normal
\Bigl\{\sqrt{n}\sigma^2{\theta t\exp(-\half\theta t)
	\over \{1-\exp(-\theta t)\}^{1/2}},1\Bigr\}. \eqno(5.3)$$
This detection power approximation is valid for small $\sigma^2$,
but illustrates well the plot's ability to detect the presence of 
a missing covariate. Note also that the whole B plot will tend to lie
above the time axis, according to these calculations, if there is
such a frailty departure from the constant hazard model. 
The A plot, on the other hand, will tend to lie above the time axis
as long as $\theta t<2$ and below the time axis when $\theta t>2$,  
as borne out by similar calculations. 
The optimal C plot uses $G_n(s)=1-\hatt\theta s$, and gives most 
weight to the smallest and then the largest time values.  

Results similar to those given here can also be reached for the 
parametric Cox regression plots of section 4. 

\bigskip
{\bf 6. General counting process models.} 
Our paper has so far been concerned with right-censored survival
times, with or without covariate information. The plots and the
results about them can be generalised with surprisingly few modifications
to much more general counting process models for life history data, 
see Hjort (1990). The necessary nonparametric and parametric machinery 
work specifically for Aalen's multiplicative intensity model,
as broadly surveyed in the ABGK book. 
Examples include models with left truncated entry times, 
competing risks, and time-inhomogeneous Markov chains, where
individuals move between states with certain hazard or transition rates.  

The practical consequence, as far as the normalised local hazard plots
are concerned, is that the formulae developed earlier are still true
as long as $Y(s)/n$ is used everywhere to estimate the asymptotic $y(s)$ 
function, where $Y(s)$ is the number of individuals at risk (for the
particular transition in question) just prior to time $s$. 
Formulae based on the $\sumjn I\{t_j\ge s\}$ identity 
are not true in this more general situation, however. 
For example, the first expression for $\hatt\Sigma_{\rm pm}$ of (2.4) 
is still valid in these more general models, but the second is not. 
We illustrate our plots in a competing risk situation in example 9.3 
and in a left-truncated entry study in example 9.4.  

An important fact about situations with several transition rates
is that the collection of Nelson--Aalen estimates are statistically
independent asymptotically. This is a fruit of the martingale theory
developed by Aalen. It also follows that the different Type B plots, say, 
for testing different parametric models for the transition rates 
under study, become approximately independent. 
This makes it easier to judge separate models for separate transition 
mechanisms in a many-state study.  

\bigskip
{\bf 7. Time-discrete models.}
There are quite similar models and methods for time-discrete survival
analysis. Although our emphasis has been clearly on the time-continuous
case we take the necessary time out to provide the most important
results for the time-discrete case. These are of interest in their
own right, since data even on time-continuous phenomena often are
recorded on a time-discrete basis, e.g.~in demography, the social
sciences, national health statistics, 
and for many censuses of central bureau of statistics type. 
The results can also be used to construct correction factors 
to the time-continuous formulae in case where data are grouped in time. 

\subsection
{\csc 7.1. General results.} 
Suppose that life-times observations are recorded only
at time points $a_0<a_1<\cdots<a_k=\tau$, where $\tau$ could be infinity.
If a distribution has point mass $f_i$ at $a_i$ then 
the hazard rate at that point is $h_i=f_i/\sum_{j\ge i}f_j$. 
Suppose a $p$-parametric model is considered of the form $h_i=h_i(\theta)$,
and let there be $\Delta N_i$ observed failures at $a_i$ among 
the $Y_i$ at risk at that point. Then the log-likelihood can be written
$\sum\{\Delta N_i\log h_i+(Y_i-\Delta N_i)\log(1-h_i)\}$,
see e.g.~Cox and Oakes (1984, chapter 3). 
The nonparametric estimate is $\hatt h_i=\Delta N_i/Y_i$.
The parametric ML estimator $\hatt\theta$ solves
$$U_n(\theta)=\sum_{a_i\le\tau}\Bigl\{{\Delta N_i\over h_i(\theta)}
        -{Y_i-\Delta N_i\over 1-h_i(\theta)}\Bigr\}h_i^*(\theta)
=\sum_{a_i\le\tau}{\Delta N_i-Y_ih_i(\theta)\over h_i(\theta)(1-h_i(\theta))}
        h_i^*(\theta)=0, \eqno(7.1) $$
where $h_i^*(\theta)={\dell\over \dell\theta}h_i(\theta)$,
a $p$-vector. One can prove that $\sqrt{n}(\hatt\theta-\theta)$ tends to a 
$\normal_p\{0,\Sigma^{-1}\}$, where 
$$\Sigma=\sum_{a_i\le\tau}{r_i\over h_i(\theta)(1-h_i(\theta))}
	h_i^*(\theta)h_i(\theta)^*
=\sum_{a_i\le\tau}r_i\psi_i(\theta)\psi_i(\theta)^\tr 
	h_i(\theta)/(1-h_i(\theta)), \eqno(7.2) $$
writing $\psi_i(\theta)={\dell\over \dell\theta}\log h_i(\theta)
=h_i^*(\theta)/h_i(\theta)$. 
Here $r_i$ is the limit in probability of $Y_i/n$, the limiting proportion
of individuals present just before time $a_i$. 
We omit the proof here, but it proceeds along lines similar to
those for the time-continuous case, as presented in Hjort (1992)
and ABGK (chapter VI), with the crucial
modification that the time-discrete martingale 
$M(t)=\sum_{a_i\le t}(\Delta N_i-Y_ih_i)$ has variance process 
$\langle M,M\rangle(t)=\sum_{a_i\le t}Y_ih_i(1-h_i)$ whereas the
time-continuous martingale $M(t)=N(t)-\int_0^tY(s)h(s)\,\d s$
has variance process 
$\langle M,M\rangle(t)=\int_0^tY(s)h(s)\,\d s$. 
Note the similarity but also the slight correction 
to the time-continuous case, cf.~(2.1)--(2.2). 

The weighted hazard difference function takes the form 
$$D_n(t)=\sqrt{n}\sum_{a_i\le t}K_i\{\Delta N_i/Y_i-h_i(\hatt\theta)\}
	=\sqrt{n}\sum_{a_i\le t}(K_i/Y_i)
	\{\Delta N_i-Y_ih_i(\hatt\theta)\}, \eqno(7.3) $$
for suitable weight function $K_i$ converging in probability to some
$k_i$. Limit theorems for time-discrete	martingales can be used to prove
$$D_n(t)\arr_dD(t)=\sum_{a_i\le t}(k_i/r_i)V_i
-\Bigl(\sum_{a_i\le t}k_ih_i^*(\theta)\Bigr)^\tr \Sigma^{-1}
\sum_{a_i\le\tau}h_i^*(\theta)V_i/\{h_i(\theta)(1-h_i(\theta))\}, $$
where the $V_i$s are independent and 
$V_i\sim\normal\{0,r_ih_i(1-h_i)\}$. The variance is 
$$\kappa(t)^2=\sum_{a_i\le t}(k_i^2/r_i)h_i(\theta)(1-h_i(\theta))
-\Bigl(\sum_{a_i\le t}k_ih_i^*(\theta)\Bigr)^\tr \Sigma^{-1} 	
\Bigl(\sum_{a_i\le t}k_ih_i^*(\theta)\Bigr), \eqno(7.4)$$
which can be estimated consistently. The end product is therefore
a NLH plot of the form $D_n(t)/\hatt\kappa(t)$, plotted for each $t=a_i$. 

As an illustration, consider the constant hazard rate model $h_i=\theta$.
Note that $\theta\in(0,1)$ in the present framework. The estimate
is $\hatt\theta=\sum_{a_i\le\tau}\Delta N_i/\sum_{a_i\le\tau}Y_i$, and 
$${\rm NLH}_B(t)={n^{-1/2}\sum_{a_i\le t}(\Delta N_i-Y_i\hatt\theta)
\over \bigl[\hatt\theta(1-\hatt\theta)v(\tau)^2\{\hatt p(t)
	-\hatt p(t)^2\}\bigr]^{1/2}}
\arr_d{W^0(p(t))\over \{p(t)-p(t)^2\}^{1/2}}\sim\normal\{0,1\} $$
for each $t=a_i$. Here $v(t)^2=\sum_{a_i\le t}r_i$ and 
$p(t)=v(t)^2/v(\tau)^2$. 

\subsection
{\csc 7.2. Observing time-continuous phenomena on a time-discrete scale.}
Suppose life-times have a continuous distribution and that 
a hazard rate model is considered of the form $h(s)=h(s,\theta)$. 
Assume however that data only are collected on a time-discrete basis,
will cells $I_i=[i,i+1)$ for $i=0,1,\ldots$. This means a time-discrete 
framework with hazards 
$$h_i(\theta)=F_\theta[i,i+1)/F_\theta[i,\infty)
	=1-\exp[-\{H(i+1,\theta)-H(i,\theta)\}], $$
and the framework above applies. Note that 
$h_i^*(\theta)=(1-h_i(\theta))\hbox{${\dell\over \dell\theta}$}
	\{H(i+1,\theta)-H(i,\theta)\}$. 
See examples 9.4 and 9.6 for illustrations. 

Let us finally mention that the above methods also lead to 
$$\Delta D_n(a_i)=\sqrt{n}K_i\{\Delta N_i/Y_i-h_i(\hatt\theta)\}
\arr_d\Delta D(a_i)\sim\normal\{0,w_i^2\}, \eqno(7.5)$$
where $w_i^2=k_i^2\{r_i^{-1}h_i(\theta)(1-h_i(\theta))
	-h_i(\theta)^2\psi_i(\theta)^\tr \Sigma^{-1}\psi_i(\theta)\}$. 
Plotting $\Delta D_n(t)$ is nugatory 
in situations with small time cells and correspondingly small values of 
$\Delta N_ih_i(\hatt\theta)$,
but in many large-scale studies in demography and social sciences
one would have substantial $Y_i$s and not too small $\Delta N_i$s.
Aalen's (1992) two examples are of this sort, for example,
and one of these are analysed as example 9.6 below. 
In such situations plotting of 
$\sqrt{n}\{\Delta N_i/Y_i-h_i(\hatt\theta)\}/\hatt w_i$ (taking $K_i=1$)
is quite informative regarding the validity of the model. 

\bigskip
{\bf 8. Concluding remarks.} 

\subsection
{\csc 8A. Maximal values of test curves.} 
Our test curves will at each given time point 
by construction stay inside $\pm1.96$ with probability 
approximately 0.95. This is a point-wise statement, and the
maximal absolute value of the random curve is 
substantially larger in distribution than the absolute
value of a standard normal. In the notation of section 2 
the maximal absolute value over $[a_1,a_2]$ tends in distribution
to $\max_{a_1\le t\le a_2}|D(t)|/\kappa(t)$, as a consequence
of process convergence. This limit distribution is quite
intricate in general, but can be handled in special cases. 

Consider a general one-parametric model $h(s,\theta)$,
and use NLH plots of Type C with $K_n(s)=n^{-1}Y(s)\psi(s,\hatt\theta)$.
Notice that the $D_n(.)$ process of (1.4) then starts and ends at zero,
by definition of the ML estimator. By covariance calculations and
Gau\ss ianeity one can show that $D_n(t)\arr_d v(\tau)W^0(p(t))$,
a scaled and time-transformed Brownian bridge, where 
$v(t)^2=\int_0^ty(s)\psi(s,\theta)^2\allowbreak h(s,\theta)\,\d s$
and $p(t)=v(t)^2/v(\tau)^2$. One can also see from (2.3) that
$\kappa(t)^2=v(\tau)^2p(t)(1-p(t))$. It follows that 
$${\rm NLH}_C(t)\arr_d{D(t)\over \kappa(t)}=
{W^0(p(t))\over \{p(t)(1-p(t))\}^{1/2}} \quad {\rm over\ }(0,\tau),$$
using convergence of stochastic processes theory. Consequently 
$$M_n=\max_{a_1\le t\le t_2}|{\rm NLH}_C(t)|\arr_d
	M=\max_{b_1\le u\le b_2}|W^0(u)|/\{u(1-u)\}^{1/2}, $$
where $b_1=p(a_1)$ and $b_2=p(a_2)$, 
involving the normalised Brownian bridge. 
An approximation to this distribution is 
$${\rm Pr}\{M\ge m\}\doteq 4\phi(m)/m+\phi(m)(m-m^{-1})\log(c_2/c_1), $$
where $c_1=b_1/(1-b_1)$ and $c_2=b_2/(1-b_2)$; see Miller and Siegmund (1982). 

If the model considered is $h(s)=\theta h_0(s)$, then 
$\psi(s,\theta)=1/\theta$ is constant and factors out, which means
that the Type C plot is in fact the same as the Type B plot,
and the above applies. Study a Type B plot for constant hazard rate,
for example, and let $a_1$ and $a_2$ be chosen as the empirical
versions of $p(a_1)=0.10$ and $p(a_2)=0.90$, where 
$p(t)=\int_0^ty(s)\,\d s/\int_0^\tau y(s)\,d s$. Then the above
shows that the maximal absolute value of the plot, over the
$[a_1,a_2]$ interval, exceeds the pointwise 1.96 limit with
probability about 0.49, and that an upper 5\% limit for this
maximum is about 3.05. 
	
Similar and in fact somewhat simpler calculations can be carried out
for the fully specified case of 3.1, where the test curve is asymptotically
distributed as $W(\kappa(t)^2)/\kappa(t)$, a time-transformed 
normalised Brownian motion process. 
Here $\kappa(t)^2=\int_0^t\{k(s)^2/y(s)\}h_0(s)\,\d s$. 

\subsection
{\csc 8B. Other empirical test curves.}
Our methods and results rely on the weak convergence results 
described in section 2. Similar results can be and have been 
reached for $\sqrt{n}\{\hatt F(t)-F(t,\hatt\theta)\}$, for example,
where $\hatt F$ is the Kaplan--Meier estimate. Test curves and 
test statistics can be constructed based on this. 
We have found it most useful to work with weighted versions 
of hazard differences instead, however, partly since 
the hazard rate quantity is more central and more easily 
generalisable to other survival data models,
and partly since results tend to be simpler.   
Our results belong to the tradition originating with 
the work by Durbin (1973) for empirical processes with estimated parameters. 
We find satisfaction in seeing practical 
and even visual uses of theoretical results in which 
``interest ... died down when the intractable 
limit processes asserted themselves'', as Pollard (1984, p.~118) remarked. 

\subsection
{\csc 8C. Testing validity over a subinterval.}
Suppose that one wishes to test a parametric model only over 
the subinterval $[a,b]$, perhaps because it is obvious that
the model cannot hold to the left of time $a$. One can then use
a C plot with $G_n$ equal to 1 on the interval and zero outside,
but this is not quite satisfactory since the plot uses $\hatt\theta$,
the ML estimate calculated from the full time interval $[0,\tau]$.
The natural remedy is to use test curves with parameter estimates
$\tilda\theta$ that only use $[a,b]$-information. 

Consider in general terms the estimator $\tilda\theta$ that solves 
$\int_0^\tau w(s)\{\log h(s,\theta)\,\d N(s)-Y(s)h(s,\theta)\,\d s\}
\allowbreak=0$.
This is an M-estimator, or a maximum weighted 
likelihood estimator, and its large-sample properties are known,
see Hjort (1985, 1992) and ABGK (chapter VI). Some work,
involving the combination of large-sample arguments of Hjort (1990) 
with such of Hjort (1992), leads to the 
following generalisation of the basic result of section 2.1:
$$\eqalign{
D_n(t)&=\sqrt{n}\int_0^tK_n(s)\{\d\hatt H(s)-h(s,\tilda\theta)\,\d s\} \cr
&\arr_d D(t)=\int_0^t\{k(s)/y(s)\}\,\d V(s) \cr
& \qquad -\Bigl(\int_0^tk(s)\psi(s,\theta)
h(s,\theta)\,\d s\Bigr)^\tr J_w^{-1}\int_0^\tau w(s)\psi(s,\theta)\,\d V(s), \cr}$$
and this zero-mean Gau\ss ian limit process has variance function
$$\eqalign{
\kappa(t)^2
&=\int_0^t(k^2/y)h\,\d s
-2\Bigl(\int_0^tk\psi h\,\d s\Bigr)^\tr J_w^{-1}\Bigl(\int_0^tkw\psi h\,\d s\Bigr) \cr
& \qquad +\Bigl(\int_0^tk\psi h\,\d s\Bigr)^\tr J_w^{-1}K_wJ_w^{-1} 
	\Bigl(\int_0^tk\psi h\,\d s\Bigr). }$$
Here $J_w=\int_0^\tau wy\psi\psi^\tr h\,\d s$ and
$K_w=\int_0^\tau w^2y\psi\psi^\tr h\,\d s$. The results previously 
used in this paper correspond to the special case 
$w(s)=1$ on the whole interval.

Let now $w=1$ on $[a,b]$ and zero outside, affecting the parameter
estimate, and let also the test weight function $K_n$ be zero outside 
the interval, affecting the test curve. Then the natural NLH curve is
$${\rm NLH}(t)=\sqrt{n}\int_a^tK_n(s)\{\d\hatt H(s)-h(s,\tilda\theta)\,\d s\}
	/\tilda\kappa(t)\quad{\rm on\ }[a,b], $$
where the denominator is the square root of a suitable estimator of
$$\int_a^t(k^2/y)h\,\d s
-\Bigl(\int_a^tk\psi h\,\d s\Bigr)^\tr \Sigma_{[a,b]}^{-1}
	\Bigl(\int_a^tk\psi h\,\d s\Bigr), $$
and where $\Sigma_{[a,b]}=\int_a^by\psi\psi^\tr h\,\d s$.  
This gives A, B and C plots for the validity of the model on $[a,b]$. 
An A plot for constant hazard rate on $[a,b]$ is for example
$\sqrt{n}\{\hatt H(t)-\hatt H(a)-\tilda\theta(t-a)\}/\tilda\kappa(t)$,
where $\tilda\theta=N[a,b]/\int_a^bY(s)\,\d s$ and 
$\tilda\kappa(t)^2=\int_a^b\{\tilda\theta/\hatt y(s)\}\,\d s
	-\tilda\theta(t-a)^2/\int_a^t\hatt y(s)\,\d s$. 
See also Hjort (1993a) for use of such interval tests to dynamic
likelihood hazard rate estimation. 

The results reported on here are also relevant to the question
of making plots with robustly estimated parameter values. 

\subsection
{\csc 8D. The first few and the last few values.}
The basic property of the plots is that ${\rm NLH}(t)$ 
is approximately distributed as a standard normal 
if the model in question is correct. This is really a large-sample
statement for each fixed $t$, as $n$ grows, and we cannot necessarily
trust the $\pm1.96$ limits to correspond precisely to 95\% coverage
probability for the smallest values of $t$. Consider the (3.2) and (3.3) 
plots for the exponential model, for example. 
Assume that there is no censoring, and let the observed life times
be ordered as $t_1<t_2<\cdots$. 
A well known transformation is to new random variables
$nt_1$, $(n-1)(t_2-t_1)$, $(n-2)(t_3-t_2)$ and so on;
these are independent and exponentially distributed with parameter $\theta$. 
Using this fact one can prove, in the framework with 
fixed $t_k$ and increasing $n$, that 
$${\rm both\ }{\rm NLH}_A(t_k){\rm\ and\ }{\rm NLH}_B(t_k)
\arr_d\sqrt{k}(1-\bar V_k)/\bar V_k^{1/2}, $$
where $\bar V_k$ is the average of independent unit exponentials 
$V_1,\ldots,V_k$. Using $\bar V_k=\chi^2_{2k}/2k$ one can 
therefore compute approximate probabilities for exceeding 1.96 in absolute
value at $t_k$, valid at least for $k$ small and $n$ large.
The first few probabilities are 0.165, 0.111, 0.092, 0.081, 0.075, 
and there is convergence to 0.05. This shows that somewhat higher 
values than under the normal can be expected for the first five or so
$t_k$s, and that not too much emphasis should be placed on 
the behaviour of the test curve here. 
(If the nonparametric $\kappa(.)$ estimators
are used instead of the parametric ones, in (3.2) and (3.3),
then the limit here is $\sqrt{k}(1-\bar V_k)$ instead.) 

The last few values have a similar behaviour, at least for
the Type B plots. With somewhat more work than for the first few 
one can prove, for the uncensored exponential case, that 
${\rm NLH}_B(t_{n-l})\arr_d\sqrt{l}(\bar W_l-1)/\bar W_l^{1/2}$,
where $\bar W_l$ is the average of i.i.d.~unit exponentials $W_1,\ldots,W_l$. 

Some remedies could be thought of in connection with these 
calculations, including a slight down-weighing at the start
and end of the plot and also skewness-reducing transformations, 
but this is not pursued here. 

\subsection
{\csc 8E. Positivity of variance estimates.} 
Some natural-looking estimates of the $\kappa^2(.)$ function of (2.3)
will turn out negative in places, even if they are perfect from
a large-sample point of view. The estimators used in this paper are 
however safe from such embarrassments. 
To see this, note first that the following holds, 
for any measure $\nu(\d s)$, provided only that
$\int_0^t vv^\tr \,\d\nu$ has an inverse: 
$$\int_0^tu^2\,\d\nu\ge\Bigl(\int_0^tuv\,\d\nu\Bigr)^\tr \Bigl(\int_0^tvv^\tr \,\d\nu
\Bigr)^{-1}\Bigl(\int_0^tuv\,\d\nu\Bigr), $$
for any functions $u\colon[0,s]\arr\RR$ and $v\colon[0,1]\arr{\RR}^p$.  
This is actually a generalisation of the Cauchy--Schwartz inequality,
and is true since the matrix
$$\Gamma=\mtrix{\int_0^tu^2\,\d\nu & \int_0^tuv^\tr \,\d\nu \cr
\int_0^tuv\,\d\nu & \int_0^tvv^\tr \,\d\nu \cr} $$
is non-negative definite, and consequently $\Gamma_{1,1}-\Gamma_{1,2}
\Gamma_{2,2}^{-1}\Gamma_{2,1}$, in usual block notation, is non-negative
definite too. And this implies 
$$\int_0^t(k^2/y)\,\d\nu\ge\Bigl(\int_0^tk\psi\,\d\nu\Bigr)^\tr 
\Bigl(\int_0^\tau y\psi\psi^\tr \,\d \nu\Bigr)^{-1}
\Bigl(\int_0^tk\psi\,\d\nu\Bigr), \quad {\rm for\ all\ }t\le\tau,$$
and again for any measure $\nu(\d s)$. In our setting $\nu(\d s)$ would 
be an estimate of $h(s,\theta)\,\d s$ and $y(s)$ would be replaced
with $Y(s)/n$. 

\subsection
{\csc 8F. Other graphical test plots.} 
One of the independent origins of the Nelson--Aalen estimator is
the 1972 paper by Nelson, advocating the sensible idea of plotting 
both the nonparametric $\hatt H(t)$ and a parametric $H(t,\hatt\theta)$
(and sometimes with ad hoc estimates for the parameters) 
in the same diagram. The difficulty of judging such pairs of estimated
cumulative hazards is that the variability differs both between the
two curves and over time. One may view our Type A plots as more worked 
out and sophisticated versions of the same idea, incorporating the correct
local stabilising scaling of the difference. 

Another sensible idea is to plot a nonparametric estimate $\hatt h(s)$
of the hazard rate itself with the parametric $h(s,\hatt\theta)$.
Versions of $\hatt h(.)$ are discussed in ABGK (chapter IV) 
and in Hjort (1991, 1993a). Some of the latter ones are inspired by the idea
of making nonparametric corrections to parametric estimates, and 
will directly or indirectly give indications of the fit of the parametric
model. Again a direct comparison can be difficult since the precision 
of the curves are quite different.  
Still other graphical model plots are discussed in 
ABGK (chapter VI). They include in particular a 
non-normalised version of our Type B plot, and plotted against $N(t)$
instead of direct time $t$. 

\subsection
{\csc 8G. The semiparametric Cox model.}
We gave methods for checking the parametric Cox model in section 4.
The possibly overused semiparametric version postulates only 
$h_j(s)=h_0(s)\exp(\beta^\tr z_j)$ with no structure imposed on the 
baseline hazard $h_0(.)$. With notation as in section~4 we have 
$\d N_j(s)=Y_j(s)\exp(\beta^\tr z_j)h_0(s)\,\d s+{\rm noise}$. 
The cumulative baseline hazard $H_0(.)$
is usually estimated using 
$$\hatt H_0(t)=\int_0^t{\d\sumjn N_j(s)\over 
	\sumjn Y_j(s)\exp(\tilda\beta^\tr z_j)\,\d s}, $$
where $\tilda\beta$ is the partial likelihood Cox estimator. 
Now consider the test function 
$$D_n(t)=n^{-1/2}\sumjn\int_0^tk_j(s)\{\d N_j(s)-Y_j(s)\exp(\tilda\beta^\tr z_j)
	\d\hatt H_0(s)\}, $$
for suitable weight functions $k_j(.)$. The choice $k_j(s)=1$ gives
simply zero, but other choices like $k_j(s)=\exp(-\tilda\beta^\tr z_j)$ 
will give curves of interest and that if the Cox model is correct should
lie around zero. We can work out the necessary asymptotic theory 
to find the limiting variance $\kappa(t)^2$ of $D_n(t)$,
after which we would have a NLH curve $D_n(t)/\hatt\kappa(t)$
to test the validity of the Cox model, but this is not pursued here. 

An increasingly popular alternative to the multiplicative Cox model
is Aalen's linear hazard nonparametric regression model, see Aalen (1989) and 
ABGK (chapter VII). Hjort (1993b) discusses also 
parametric versions of this model, and the validity of such can again
be assessed using appropriate test curve constructions of the NLH variety. 
We have worked out the necessary theory but this will be presented
elsewhere. 


\bigskip
{\bf 9. Examples and illustrations.} 

\subsection
{\csc Example 9.1. Testing a constant hazard.}
\fermat{figures are placed at the end of our report}Figure 9.1a displays NLH plots of Type A and Type B for 
100 simulated unit exponentials (i.e.~having constant hazard rate 1). 
Figure 9.1b presents A and B curves in a situation with 
100 simulated unit exponentials but censored with another 
100 simulated unit exponentials. The plots behave as expected,
cf.~also Remark 8A. 

{\medskip\narrower\noindent\sl
{\csc Figure 9.1.} 
NLH plots of Type A (line) and Type B (dashed) are shown in 9.1a 
for 100 simulated unit exponentials. 
9.1b shows A and B curves for 100 unit exponentials in a situation
with about 50\% censoring (the censoring variables are another set
of 100 unit exponentials).
\smallskip}
  
\subsection
{\csc Example 9.2. Testing a Weibull.} 
100 variables simulated from the Weibull (10,1.3) distribution,
i.e.~with hazard rate $13s^{0.3}$,  
gave rise to the plots of 9.2a and 9.2b. 
Type A plots for the exponential and the Weibull are shown 
in 9.2a and Type B plots in 9.2b. The curves for the exponential model
clearly wanders away from the acceptable band while the Weibull curves
stay within. Note also that the plots indicate a Weibull shape parameter
greater than 1, in view of remarks made in section 5. 
A Weibull (10,0.7) distribution would for example 
give exponential model plots that first increased and then decreased. 

{\medskip\narrower\noindent\sl
{\csc Figure 9.2.} 
NLH plots of Type A are shown for the exponential model (line) 
and Weibull model (dashed) are shown in 9.2a 
for 100 simulated Weibull (10,1.3) variables. 
NLH plots of Type B for the same data are shown in 9.2b. 
\smallskip}


\subsection
{\csc Example 9.3. IUD expulsion and removal data.}
These data are from Peterson (1975) and are the experience 
of a sample of 100 women using an experimental intrauterine 
contraceptive device (IUD). 
There are several competing risks in this experiment, 
and following Aalen's 1982 analysis we have focused on two: 
unplanned removal and expulsion of the IUD. Most of the removals were 
planned, leading to heavy censoring. A plot of the Nelson--Aalen estimator 
of the two cumulative hazards are given in figure 9.3a. 
The cumulative hazard for unplanned removal appears to be linear, 
and that for expulsion appears to increase logarithmically. 
This suggests an exponential model for
removals and a simple frailty model for expulsions.
The NLH plots for unplanned removal are given in figure 9.3b and show good
agreement with the model. Figure 9.3c, comparing the IUD expulsions with
the simple frailty model, shows  close agreement. The dramatic step in the
model at a time of one year is due to the planned removal of IUDs (censoring)
in a large proportion of the remaining women at this time.
Our conclusions agree with more informal analysis by Aalen (1982). 

{\medskip\narrower\noindent\sl
{\csc Figure 9.3.} 
Nelson--Aalen estimators for the cumulative hazard rates for 
expulsion (dotted line) and unplanned removal (line) are shown in 9.3a. 
NLH plots for constant hazard rate are given in 9.3b for unplanned removal
and for the simple frailty model of 3.5 in 9.3c for expulsion. 
As in figures 9.1 and 9.2 the A plots are with a line and the B plots 
are dashed. 
In both cases there is close agreement with the model.
\smallskip}

\subsection
{\csc Example 9.4. Gompertz mortality rates for Fyn diabetics.}
ABGK (example I.3.2 and later on) discuss several aspects of data
that have been obtained by Green et al.~(1981) on the mortality 
for 716 women and 783 men suffering from insulin-dependent diabetes
mellitus in the Danish county of Fyn. We have got the full data set from
Andersen and Borgan (personal communication). The life-times are
partly right-censored since the study was finished 1 January 1982, 
and also partly left-truncated, since only persons alive by 1 July 1973
could enter the study. Although the traditional framework of sections 2 and 3
is too narrow our methods still work, as explained in section 6,
with $Y(s)$ of the form $\sumjn I\{t_j\ge s>v_j\}$, where $v_j$ denotes
entry age and $t_j$ is exit age (age of death if death occurred 
before 01.01.1982 and age at this date otherwise). 

ABGK consider the women group of this example in some depth 
in their chapter VI, and we shall complement their analysis by 
studying the fit of the time-continuous Gompertz model 
$h(s)=\theta\exp(\beta s)$ for the mortality rate. 
The data are only given in whole years, and there are ties, 
so we shall in fact use the time-discrete machinery outlined in section 7.
There are $Y_i$ women alive and diagnosed with diabetes 
when entering year interval $[i,i+1)$, and $\Delta N_i$ 
of these die during this year, where $i=1,2,\ldots,98$. 
This framework is slightly more precise, presumably, given the grouped
data, although the time-continuous apparatus also is acceptable 
here as an approximation, since the time intervals are relatively short 
(ABGK use time-continuous techniques).  
The hazard for the $i$th year interval is 
$$h_i(\theta,\beta)=F[i,i+1)/F[i,\infty)=1-\exp\{-H[i,i+1)\}
=1-\exp\bigl[-(\theta/\beta)(e^{(i+1)\beta}-e^{i\beta})\bigr]. \eqno(9.1)$$
ML estimates 
0.957/$10^3$ (0.317/$10^3$) and 61.110/$10^3$ (4.668/$10^3$) 
for $\theta$ and $\beta$
were found by a Newton--Raphson algorithm
(estimated standard deviations in parentheses). ABGK use 
an alternative parametrisation $bc^{s-50}$ and finds (p.~413) 
0.0199 and 1.066 for $b$ and $c$ using a likelihood appropriate
to time-continuous data; our values give instead 0.0203 and 1.063. 
Using our ML estimates we can display nonparametric versus parametric
estimates of cumulative hazards, and thereby produce a close relative of 
figure IV.3.5 in ABGK, but perhaps with even better fit 
for age $\ge80$ years.  

{\medskip\narrower\noindent\sl
{\csc Figure 9.4.} 
Time-discrete NLH plots of Type A (connected dots) and Type B 
(connected plusses) for the Gompertz model are shown for
the group of 716 women in 9.4a and for the 783 men in 9.4b. 
In both cases there is close agreement with the model. 
NLH plots to compare the mortality rate for men with diabetes 
with the estimated Gompertz mortality rate for women with diabetes 
are given in 9.4c, and shows that the mortality rate for men is higher.
\smallskip}

The A and B plots to assess the validity of the Gompertz model,
or strictly speaking rather the inherited model (9.1) for 
yearly hazards, are shown in figure 9.4a, 
and are a convincing show of support for the model.
We note that ABGK were able to detect a certain unexplained 
departure from the Gompertz model using a Khmaladze test
(see their example VI.3.9). This difference in opinion might
partly stem from their use of time-continuous machinery and 
perhaps partly from their use of a $\tau$ value at the very end of the time 
scale (note their discussion of this ``very delicate matter'' on p.~466). 
To indicate the time-discreteness the A plot is shown with yearly dots. 
A similar analysis was carried out for the men group (not similarly analysed
in ABGK). ML estimates (again using the time-discrete likelihood) are 
1.097/$10^3$ (0.314/$10^3$) and 64.809/$10^3$ (4.390/$10^3$) 
for $\theta$ and $\beta$. 
This translates to 0.028 and 1.067 for $b$ and $c$. 
The A and B plots for this group are given in 9.4b. 
We conclude that both men and women of Fyn look perfectly Gompertzian to us. 
To show that men and women are different we have also included 
a Type A plot to assess the hypothesis that the 
men have the hazard rates $h_i=h_{i,0}$ as specified by (9.1) with 
parameter estimates from the women stratum. In this fully specified
case the variance expression (7.4) simplifies by losing its second term,
cf.~3.1. 

\subsection
{\csc Example 9.5. Parametric Cox regression for Danish melanoma survival data.}
This data set, though with fewer variables, was analysed extensively in 
ABGK. The full set has been given us by Andersen and
Borgan (personal communication). They relate to 205 patients
who had a particular operation to remove malignant melanoma (a form of skin
cancer) in Odense, Denmark in the period 1962--77. Various risk factors were
recorded at the time of the operation. The patients were observed
until 1977, and the time (in days) until they died of melanoma was 
recorded. Patients dead of other causes or still alive in 1977 are 
treated as censored observations.  

One of the analyses performed by ABGK was a semiparametric 
Cox regression. The estimated baseline cumulative hazard appeared fairly 
nearly linear, leading us to select these data for a parametric analysis 
with constant baseline hazard. We used the `Poisson error trick' 
of Aitkin and Clayton (1980) to fit the model as a generalised linear model 
with Poisson error and log link. ABGK used thickness of the 
tumour (in mm) and sex of the patient as covariates, 
and our data set has a further three: the layer of skin to which the tumour 
penetrated (coded 1 to 4), presence of ulceration on the tumour, 
and presence of a certain type of cell (epithelioide cells) in the tumour. 
Men had a higher hazard than women, and the actions of the other
variables were all in the directions which would be expected. 
In the parlance of generalised linear models 
the deviance of the resulting model was 187.1 on 198 degrees 
of freedom (null deviance was 232.1 on 204 df), giving an acceptable fit.
A measure of the effect of a variable $z$ is the range of values of  
$\beta z$, which measures how much deviation from  uniform hazard 
is attributable to this variable. The variables with the highest values 
of $\Var(\beta_iz_i)$, and so associated with the greatest variation 
in hazard, were `presence of ulceration' and `skin layer penetrated'.

The NLH plots of this exponential regression model show the 
expected relationship between the exponential and regression model plots.  
In the type B plot, figure 9.5b, 
the curve for the exponential model lies above that 
for the regression model, and in the Type A plot, figure 9.5a, 
the curves cross as expected, cf.~remarks at the end of section 5. 
The plots do not, however, indicate a better fit from the regression model. 
This calls into question the assumption of constant baseline hazard 
over the whole time period. Looking at the
data we find that are no deaths from melanoma after about 9 years, 
and fewer than would be expected in the first two years. 
The sparse nature of the data after 9 years means 
that the estimated variance is very high and
so the NLH-plots only show the initial dip in hazard.
The nonparametric variance estimates are not as sensitive to
the spacing of points and should work better at the end of the
time interval, and in fact figure 9.5c shows that they allow 
the exponential model to be rejected by a Type A plot. 

Figure 9.5d displays an estimate of the baseline hazard for a 
semiparametric Cox model fitted to these data. 
The hazard is estimated by kernel smoothing 
$\int b^{-1}K(h^{-1}(s-t))\,\d\hatt H_0(t)$
of the estimated cumulative hazard given in 8G. 
The kernel function chosen is 
$K(z)={15\over 8}(1-8z^2+16z^4)$ on $[-\half,\half]$,
since it is the simplest one obeying the natural requirements
of being a symmetric unimodal probability density with existing
derivatives zero at $\pm\half$ (the hazard estimate will not 
have a continuous derivative without the last requirement,
which is why we avoided the Yepanechnikov kernel). 
We have worked out some theory for cross validation and 
for adaptive smoothing and also 
experimented with orthogonal expansion estimators. 
A uniform bandwidth value of 4 years was found to be satisfactory. 
An estimate for times near zero was achieved by augmenting the 
observed points $t_1,t_2,\ldots$ with `reflected' points 
$-t_1,-t_2,\ldots$, following a boundary technique for density
estimation, see for example Scott (1992, section 6.2.3.5).
This allows estimation of the baseline hazard down to $s=0$ 
but constrains it to have zero derivative at $s=0$. 

The plots we have given show that the constant baseline hazard suggested by
the cumulative hazard plot cannot be justified under closer examination. 
We also mention that ABGK (example VII.3.1) 
as well as Murphy (1993) have pointed to some problems with the 
proportional hazards assumptions for these data. 
It should be noted, however, that a semiparametric Cox model 
that was fitted in order to produce figure 9.5d 
gave almost identical coefficient estimates and overall significance level,
suggesting that the conclusions in this case are not very sensitive to 
assumptions made in the analysis.



{\medskip\narrower\noindent\sl
{\csc Figure 9.5.} 
NLH-plots of Type A and B are in respectively 9.5a and 9.5b 
for the parametric Cox regression model (dashed lines) 
and the homogeneous exponential model (solid line). 
In 9.5c an A plot for the homogeneous exponential model is given,
suggesting that the true hazard is lower at the beginning and at the end
of the time interval. 
Figure 9.5d gives a kernel smoothed estimate of the 
baseline hazard in a semiparametric Cox model for the melanoma data.  
\smallskip}

\subsection
{\csc Example 9.6. Time-discrete four-parameter heterogeneity model
for the time to next birth.} 
Data have been extracted from the Norwegian Medical Birth Registry 
by O.~Aalen and B.~Sandstad to find the time to next birth 
for young women experiencing a stillbirth. 
The data set consists of all the 451 Norwegian women who had their first 
birth during 1967 to 1971, who were at the time of this birth below
25 years of age and married, and for whom the child was stillborn. 
We have got the data from Aalen (personal communication). 
Aalen (1992) discusses this data set and fits a four-parameter frailty model:
woman $j$ is thought of as having `hazard' or intensity rate 
$Z_j\lambda(t)$, where $\lambda(t)$ is a Weibull starting after nine
months, while the distribution of the $Z_j$s among the women 
is thought to have a certain compound Poisson distribution.
The result is as in (3.12), with 
$$\lambda(t)=a(t-9/12)^k \quad{\rm and}\quad
\Lambda(t)=a(t-9/12)^{k+1}/(k+1), \quad {\rm for\ }t\ge9/12. \eqno(9.2)$$
The data are only collected time-discretely, however, so we cannot
produce NLH plots to test (9.2) directly. There are 24 time intervals
of variable lengths, starting with $(9/12,10/12)$, $(10/12,11/12)$ 
and ending with $(12,13)$, $(13,15)$ (years). 
Letting the $i$th time interval be $(t_{i,l},t_{i,r})$ the number $Y_i$
of women having not yet given birth after the stillbirth when entering
this time interval is known, as is the number $\Delta N_i$ of these
that then give birth within this interval. The hazard rate for 
the $i$th interval is 
$$h_i=h_i(a,k,\alpha,\delta)=1-\exp[-\{H(t_{i,r},a,k,\alpha,\delta)
	-H(t_{i,l},a,k,\alpha,\delta)\}], \eqno(9.3)$$
with the $H(.)$ function as in (3.12), and the log-likelihood is 
$\sum_{i=1}^{24}\{\Delta N_i\log h_i+(Y_i-\Delta N_i)\log(1-h_i)\}$. 
We wrote a Newton--Raphson programme in \Splus{} to find ML estimates
5.141 (1.600) for $a$, 
1.152 (0.193) for $k$,
1.305 (0.114) for $\alpha$, and
1.551 (0.315) for $\delta$. 
This required lengthy partial derivatives calculations. 
The numbers in parentheses are the estimated standard errors,
obtained from $\Sigma^{-1}/n$ and (7.2). The $a$ value is incorrectly
given in Aalen (1992), 
but his plots showing a quite good agreement 
between model and data are correct. 
(He also estimated the standard errors differently.) 
In figure 9.6 the time-discrete normalised hazard difference 
$\sqrt{n}\{\Delta N_i/Y_i-h_i(\hatt a,\hatt k,\hatt\alpha,\hatt\delta)\}/
\hatt w_i$ is plotted against the midpoints of the time intervals, 
see (7.5). The plot indicates a very good fit.
Similarly the three-parameter model where $k=1$, corresponding
to a linearly increasing individual intensity rate, would also 
give a quite accurate fit. 

Of course these three- and four-parameter classes are quite rich and one 
should not overstate this particular probabilistical 
explanation of ${\cal N}${\hskip-1.5pt}ature. 
It is nevertheless an attractive model 
with a natural socio-biological interpretation and an impressive fit. 
Note that each woman has a steadily increasing intensity rate for 
giving birth a second time (up to 40 years of age). 
A feature of the model is also that a certain proportion of the women, 
namely $\exp(-{\alpha\over \alpha-1}\delta)$ and here estimated to be 
6.4\%, will never have the second birth. 

{\medskip\narrower\noindent\sl
{\csc Figure 9.6.} 
Time-discrete normalised hazard difference plot assessing the 
compound Poisson heterogeneity model for the time to next birth following
stillbirth for 451 young Norwegian women. 
\smallskip}

\bigskip
{\bf Appendix: \Splus{} procedures.} 
A package of \Splus{} procedures has been produced to produce 
the Type A and Type B plots and to perform the necessary parameter 
estimation for the exponential, Weibull, Gompertz, simple frailty 
and exponential (Cox) regression models. These procedures were used 
to produce the plots in this paper, and
are available by electronic mail from either of the authors
upon receipt of an ethnic postcard.  

The package consists of a single function for the user to call that
then sorts the data and calls separate procedures to compute the appropriate
maximum likelihood parameter estimates and calculate and draw all 
the requested plots. The parameter estimates and approximate 
standard deviations are also reported.

These procedures calculate the normalised local hazards only at times when
cases are observed, and interpolates linearly between these points. 
This does not change the asymptotic behaviour of the plots, but  
saves computation and produces smoother pictures for small samples.
The parameter estimates use a variety of methods. There is an explicit formula
for the exponential model. The exponential regression model can be rewritten 
as a generalised linear model using the Poisson error trick explained
for example in Aitkin and Clayton (1980), and is then fitted 
using the {\tt glm} procedure in \Splus. The other models are 
fitted by direct numerical maximisation of the log-likelihood. 
The variance-covariance matrix of the parameter estimates is approximated by
$\Sigma^{-1}/n$, with the same matrix $\Sigma^{-1}$ as is 
used in calculating the plots.
The calculations for the plots are carried out entirely within \Splus, 
rather than using C or FORTRAN routines. While \Splus{} procedures are
easier to write, debug and extend than C or FORTRAN, they can be much slower;
a plot for 500 cases can take a couple of minutes. 

It should be a straightforward exercise to extend the procedures to compute
different NLH plots, once the correct formulae are derived. The only 
significant complications arise from the structure of \Splus. 
While there are efficient commands for manipulating matrices, 
vectors and lists as single objects, loops to manipulate components 
of these structures can run very slowly. For this reason, 
the integrals required in the plots are calculated by first working 
out the increments between successive time points and then taking 
cumulative sums of these increments. If other systems were used instead of 
\Splus, it would probably be easiest to use the
simpler finite-sum expressions.

\bigskip 
{\bf Acknowledgements.} 
We are grateful to Odd Aalen, \O rnulf Borgan and Per Kragh Andersen 
for giving us copies of various data sets with necessary explanations, 
for permission to use them, and for their interest. 
This work was carried out while the first author was visiting Oxford
with a grant from the Royal Norwegian Research Council.

\def\ref#1{{\noindent\hangafter=1\hangindent=20pt
  #1\smallskip}}  

\bigskip  
\centerline{\bf References}

\smallskip
\parindent0pt
\baselineskip11pt
\parskip3pt

\ref{%
Aitkin, M.~and Clayton, D.G. (1980).
The fitting of exponential, Weibull and extreme value distributions
to complex censored survival data using {\csc glim}. 
{\sl Applied Statistics}~{\bf 29}, 156--163. } 

\ref{%
Andersen, P.K., Borgan, \O., Gill, R.D., and Keiding, N. (1993).
{\sl Statistical Models Based on Counting Processes.}
Springer-Verlag, New York. }

\ref{%
Arulchelvam, M. (1992).
Some topics pertaining to parametric survival data models. 
Cand.~Scient.~thesis, Department of Mathematics and Statistics, 
University of Oslo.}

\ref{%
Cox, D.R.~and Oakes, D. (1984).
{\sl Analysis of Survival Data.}
Chapman and Hall, London.}

\ref{%
Durbin (1973). 
{\sl Distribution Theory for Tests Based on the Sample Distribution 
Function.} SIAM, Philadelphia. }


\ref{%
Green, A., Hauge, M., Holm, N.V.~and Rasch, L.L. (1981).
Epidemiological studies of diabetes mellitus in Denmark. 
II. A prevalence study based on insulin prescriptions. 
{\sl Diabetologia} {\bf 20}, 468--470.}

\ref{%
Hjort, N.L. (1985).
Contribution to the discussion of Andersen and Borgan's
`Counting process models for life history data: a review'. 
{\sl Scandinavian Journal of Statistics}~{\bf 12}, 141--150. }
 
\ref{%
Hjort, N.L. (1990).
Goodness of fit tests in models for life history data based on 
cumulative hazard rates. 
{\sl Annals of Statistics}~{\bf 18}, 1221--1258. }

\ref{%
Hjort, N.L. (1991).
Semiparametric estimation of parametric hazard rates.
In {\sl Survival Analysis: State of the Art}, 
Kluwer, Dordrecht, pp.~211--236. 
Proceedings of the {\sl NATO Advanced Study Workshop on Survival
Analysis and Related Topics}, Columbus, Ohio, eds.~P.S. Goel and J.P.~Klein.} 

\ref{%
Hjort, N.L. (1992). 
On inference in parametric survival data models. 
{\sl International Statistical Review} {\bf 60}, 355--387.}

\ref{%
Hjort, N.L. (1993a).
Dynamic likelihood hazard rate estimation.
Submitted for publication.}

\ref{%
Hjort, N.L. (1993b).
Efficiency of three estimators in Aalen's linear hazard rate regression model.
Submitted for publication.}

\ref{%
Hjort, N.L.~and Lumley, T. (1993).
Normalised Local Hazardplots: a \Splus{} package.
Freely available by electronic mail upon receipt of an ethnic postcard
to either of the authors.}

\ref{%
Hjort, N.L.~and Pollard, D.B. (1993).
Asymptotics for minimisers of convex processes.
Submitted for publication.} 

\ref{%
Koning, A. (1991).
Stochastic integrals and goodness-of-fit tests.
Proefschrift ter verkrijging van de graad van doctor, 
Universiteit Twente.} 



\ref{%
Miller, R.G.~and Siegmund, D. (1982).
Maximally selected chi-square statistics.
{\sl Biometrics} {\bf 38}, 1011--1016. }

\ref{%
Murphy, S.A. (1993).
Testing for a time dependent coefficient in Cox's regression model.
{\sl Scandinavian Journal of Statistics}~{\bf 20}, 35--50.}

\ref{%
Nelson, W. (1972).
Theory and applications of hazard plotting for censored failure data.
{\sl Technometrics} {\bf 14}, 945--965.}

\ref{%
Peterson, A. V. (1975). 
Nonparametric estimation in the competing risks problem. 
PhD thesis, Department of Statistics, Stanford University.}

\ref{%
Pollard, D.B. (1984). 
{\sl Convergence of Stochastic Processes.}
Springer-Verlag, New York.}


\ref{%
Scott, D.W. (1992).
{\sl Multivariate Density Estimation: 
Theory, Practice, and Visualization.}
Wiley, New York.}

\ref{%
Aalen, O.O. (1982). 
Practical applications of the non-parametric statistical theory
for counting processes. Statistical Research Report, 
Department of Mathematics and Statistics, University of Oslo.}

\ref{%
Aalen, O.O. (1989).
A linear regression model for the analysis of life times.
{\sl Statistics in Medicine} {\bf 8}, 907--925.} 

\ref{%
Aalen, O.O. (1992).
Modelling heterogeneity in survival analysis by the compound Poisson
distribution. {\sl Annals of Applied Probability}~{\bf 2}, 951--972.} 

\def\quotationone{\smallrm }
\def\quotationtwo{\smallrm }

\vfill\eject


\centerline{\includegraphics[scale=0.55]{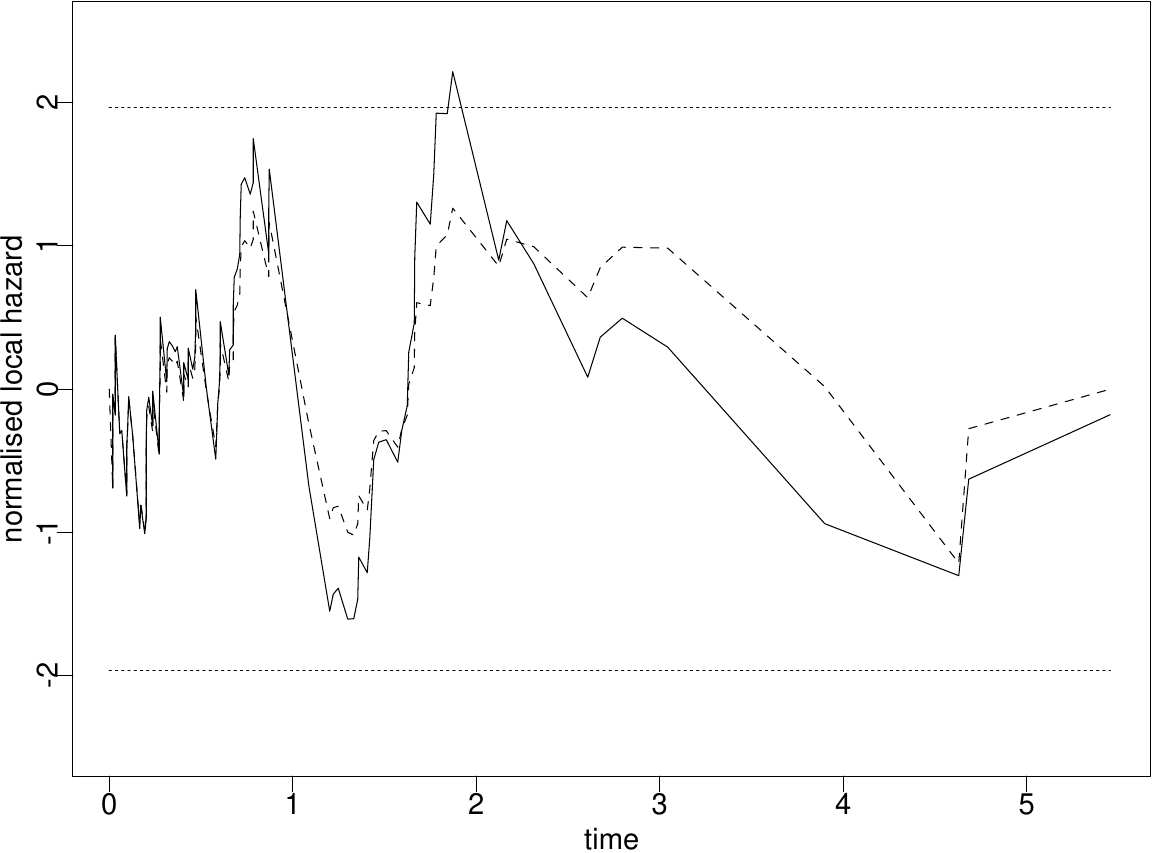}}

\bigskip

\centerline{\includegraphics[scale=0.55]{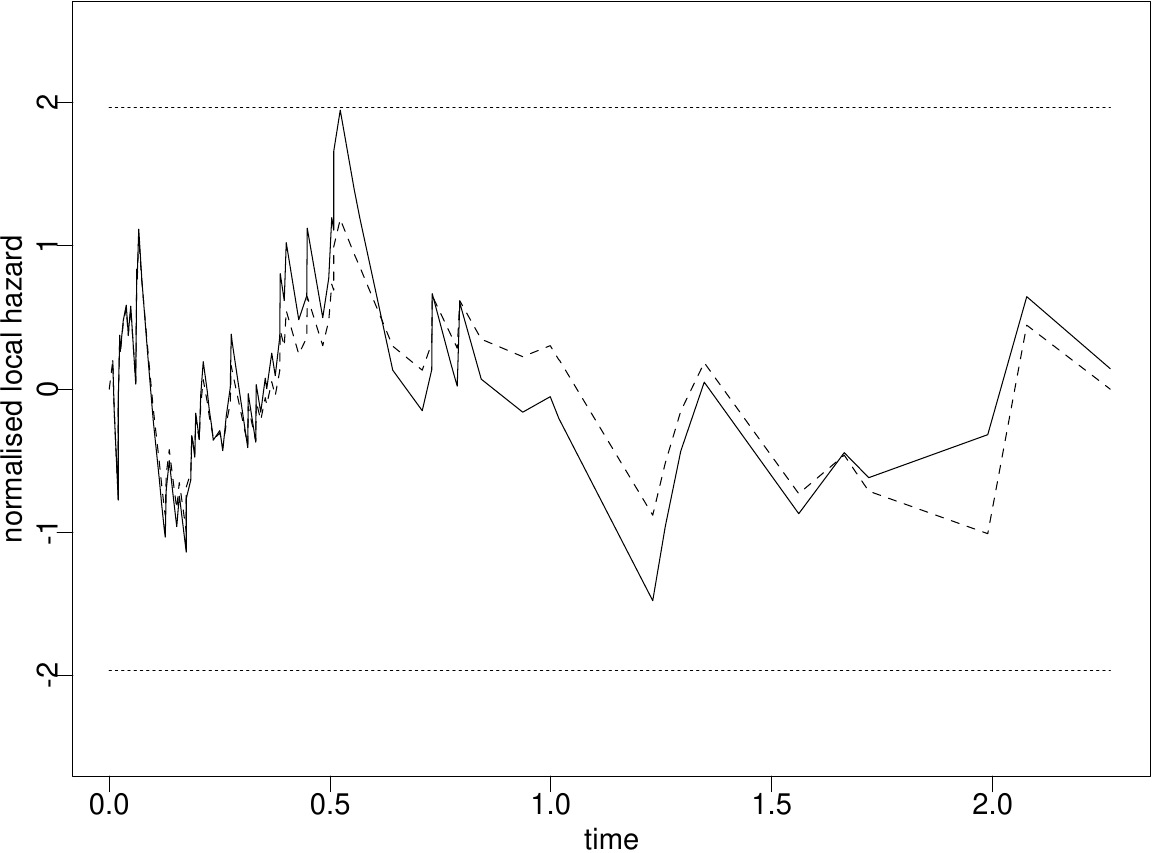}}

{\bigskip\narrower\noindent\sl
{\csc Figure 9.1.} 
NLH plots of Type A (line) and Type B (dashed) are shown in 9.1a 
for 100 simulated unit exponentials. 
9.1b shows A and B curves for 100 unit exponentials in a situation
with about 50\% censoring (the censoring variables are another set
of 100 unit exponentials).
\smallskip}

\vfill\eject 


\centerline{\includegraphics[scale=0.55]{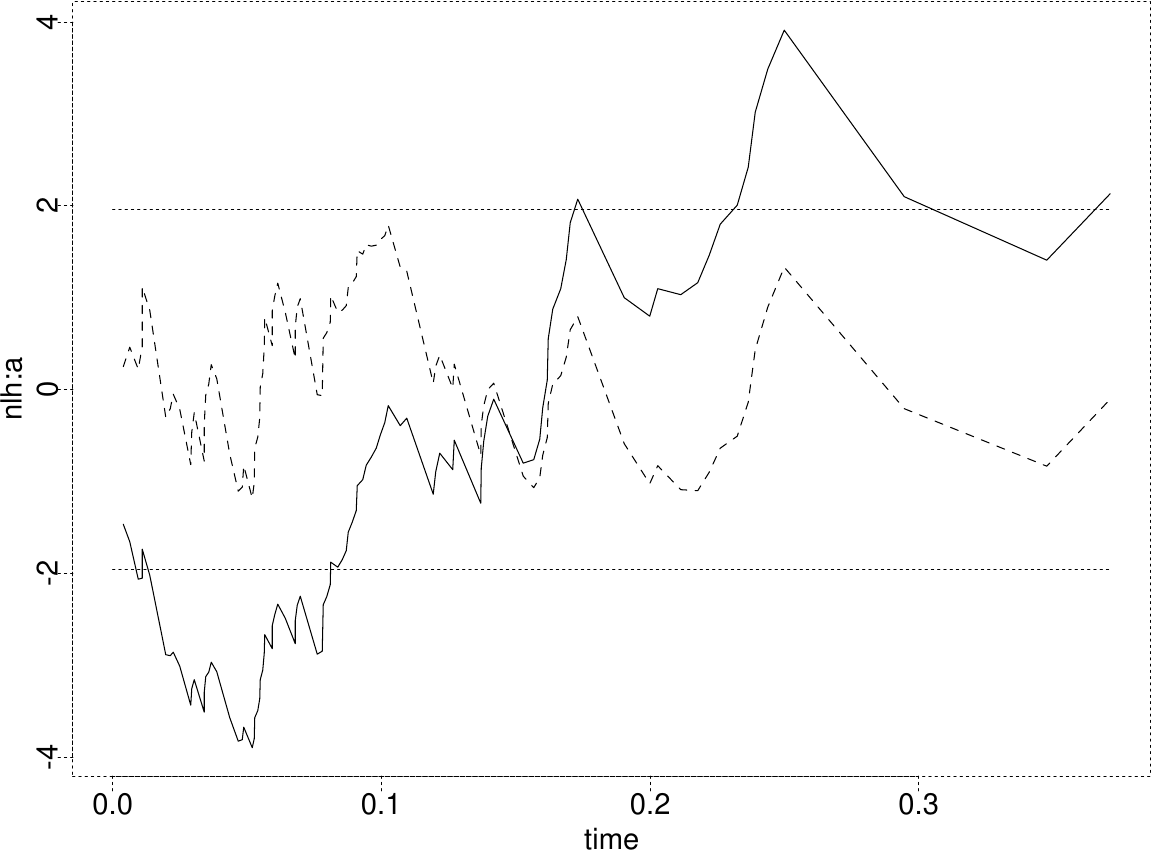}}

\bigskip

\centerline{\includegraphics[scale=0.55]{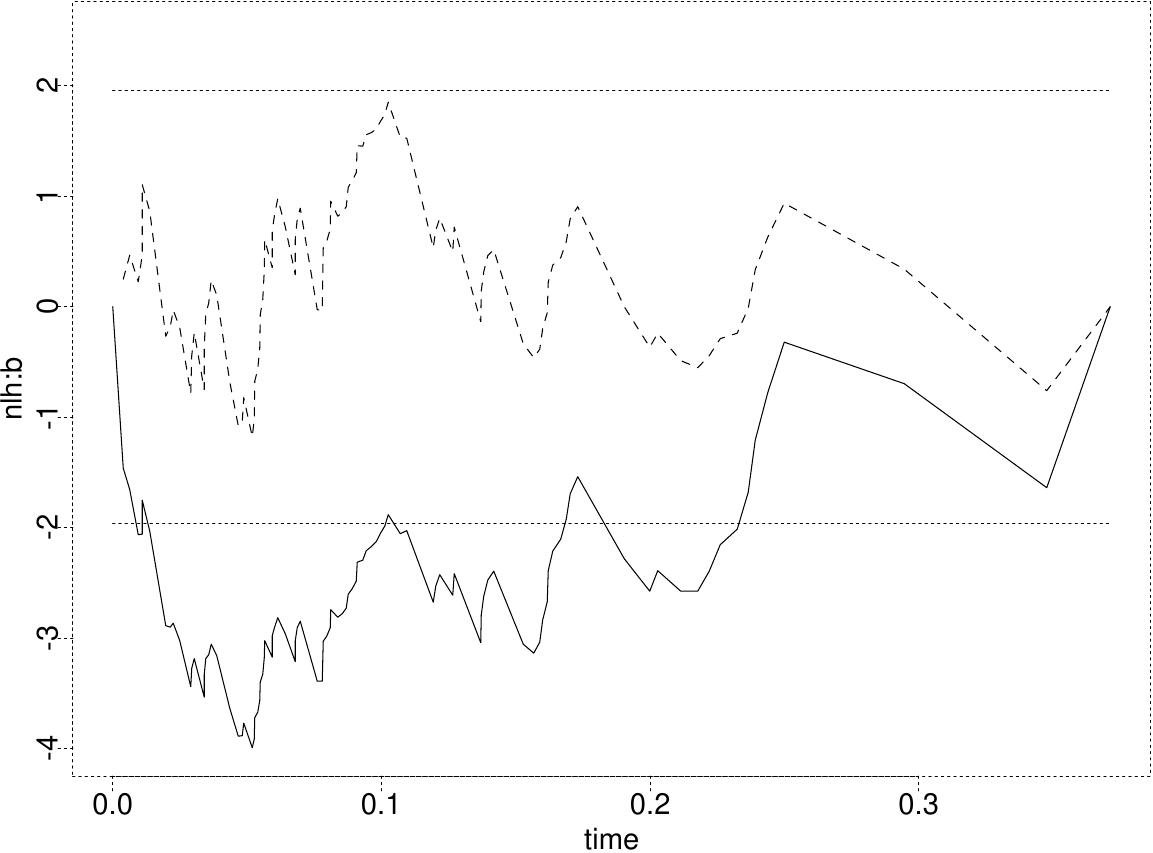}}

{\bigskip\narrower\noindent\sl
{\csc Figure 9.2.} 
NLH plots of Type A are shown for the exponential model (line) 
and Weibull model (dashed) are shown in 9.2a 
for 100 simulated Weibull (10,1.3) variables. 
NLH plots of Type B for the same data are shown in 9.2b. 
\smallskip}

\vfill\eject 


\centerline{\includegraphics[scale=0.55]{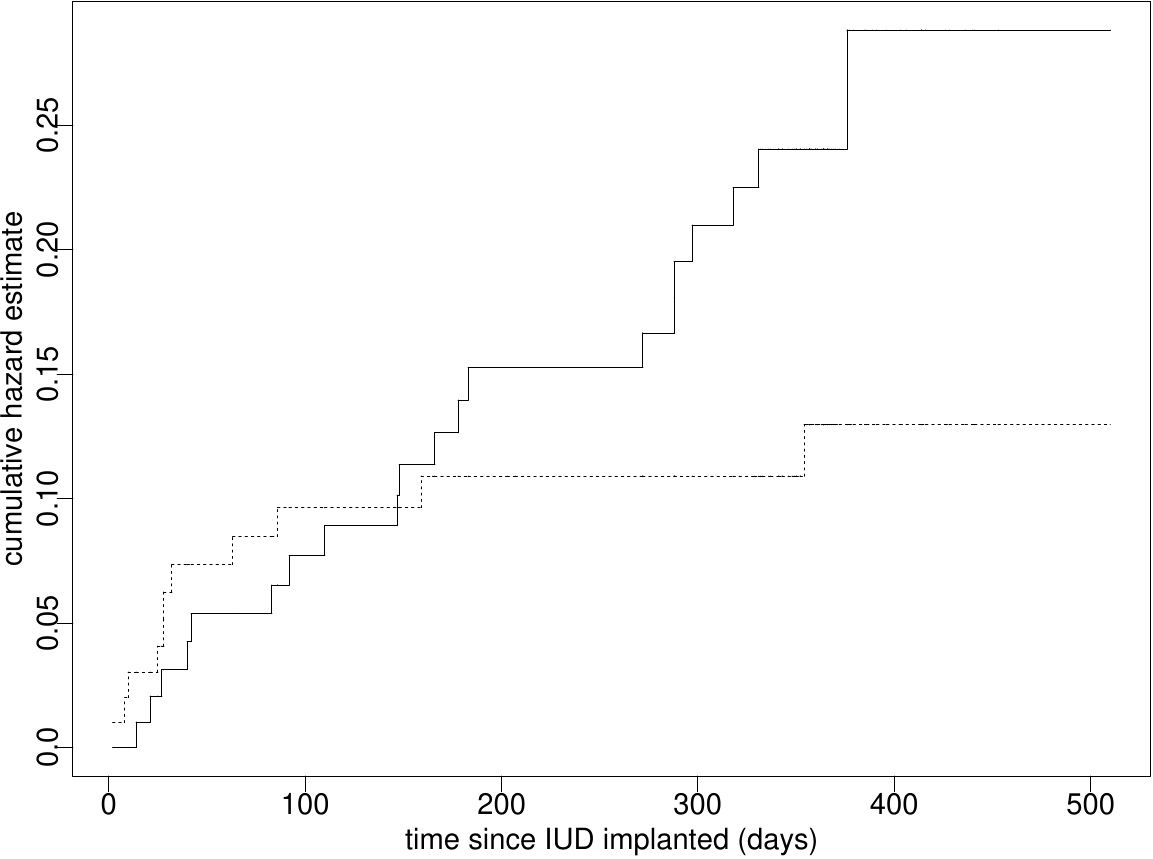}}

\bigskip

\centerline{\includegraphics[scale=0.55]{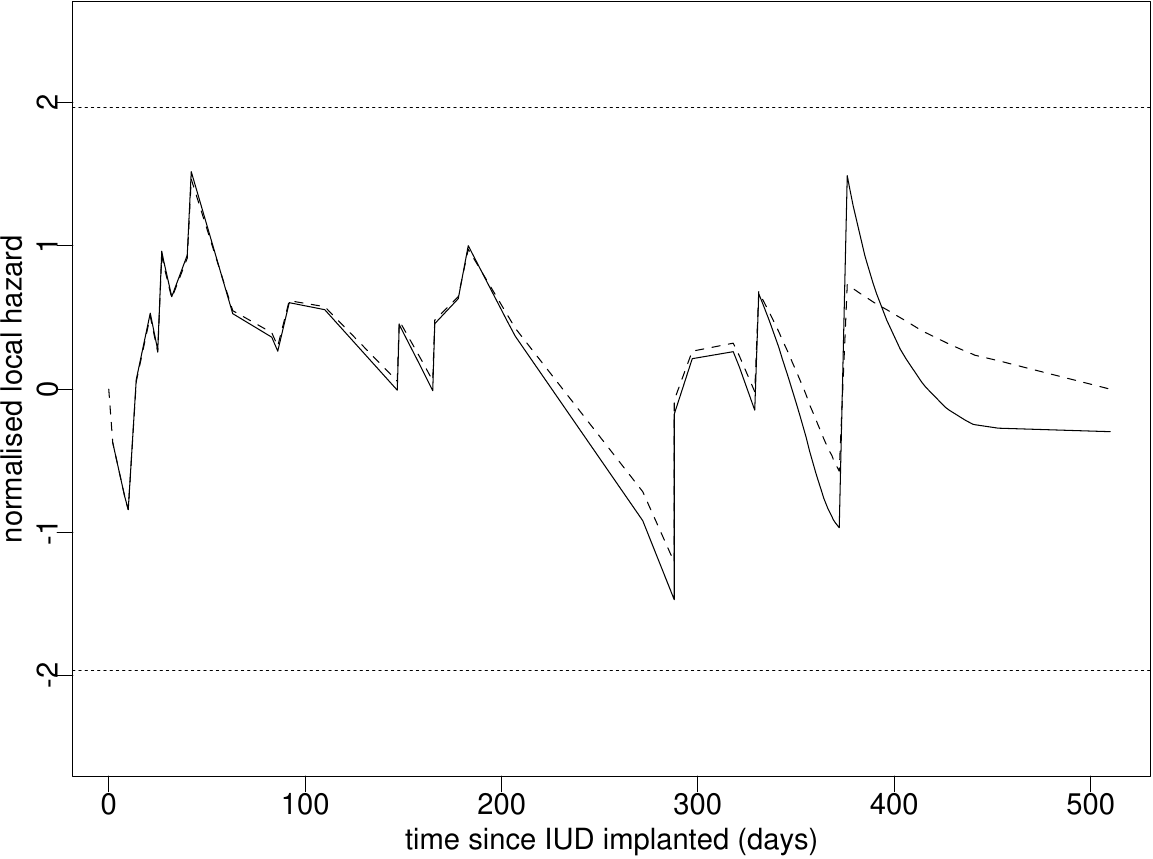}}

\bigskip
\centerline{\includegraphics[scale=0.55]{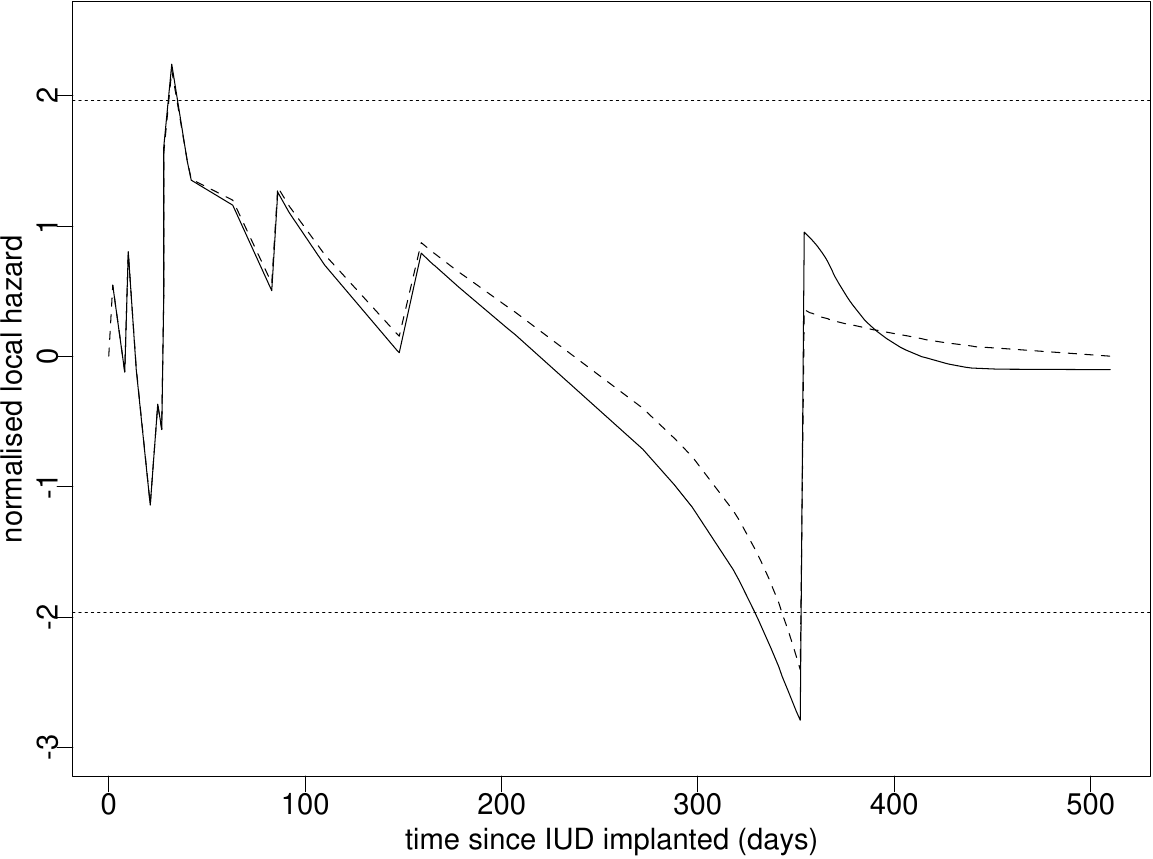}}

{\bigskip\narrower\noindent\sl
{\csc Figure 9.3.} 
Nelson--Aalen estimators for the cumulative hazard rates for 
expulsion (dotted line) and unplanned removal (line) are shown in 9.3a. 
NLH plots for constant hazard rate are given in 9.3b for unplanned removal
and for the simple frailty model of 3.5 in 9.3c for expulsion. 
As in figures 9.1 and 9.2 the A plots are with a line and the B plots 
are dashed. 
In both cases there is close agreement with the model.
\smallskip}

\vfill\eject 


\centerline{\includegraphics[scale=0.55]{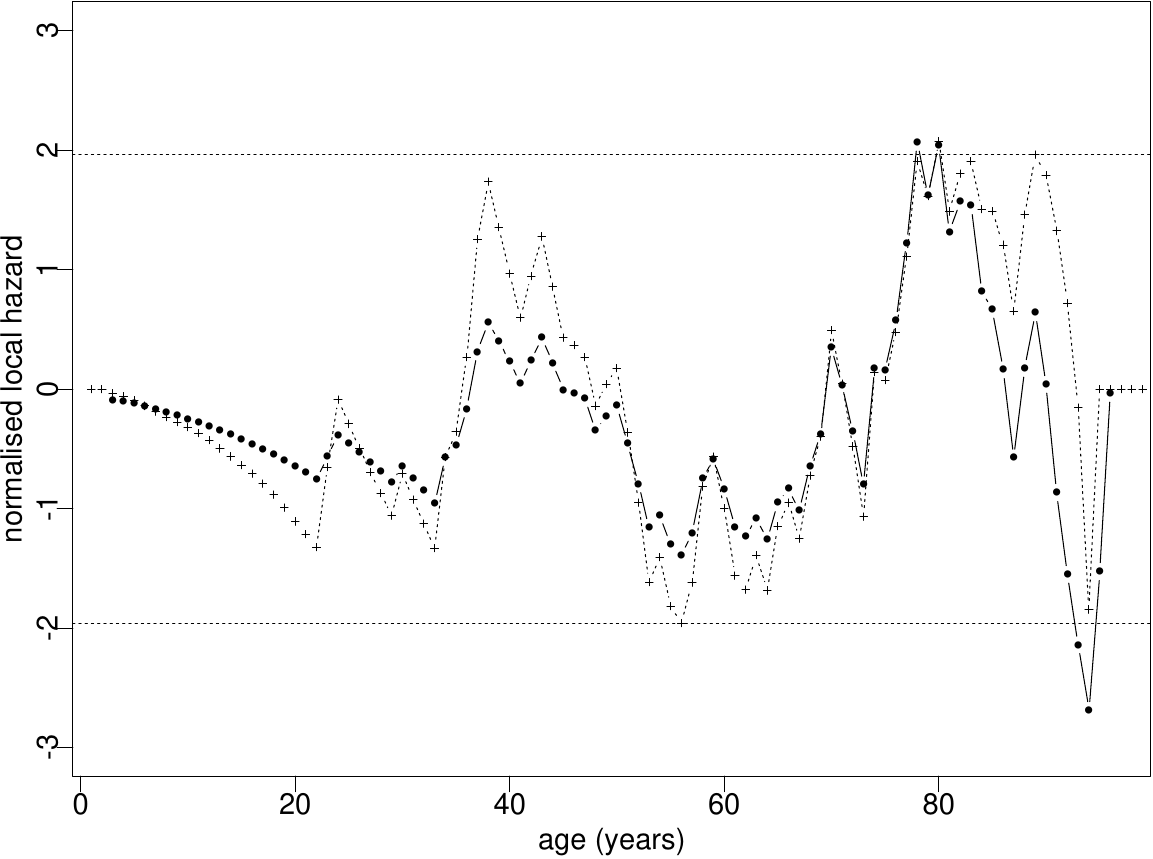}}

\bigskip

\centerline{\includegraphics[scale=0.55]{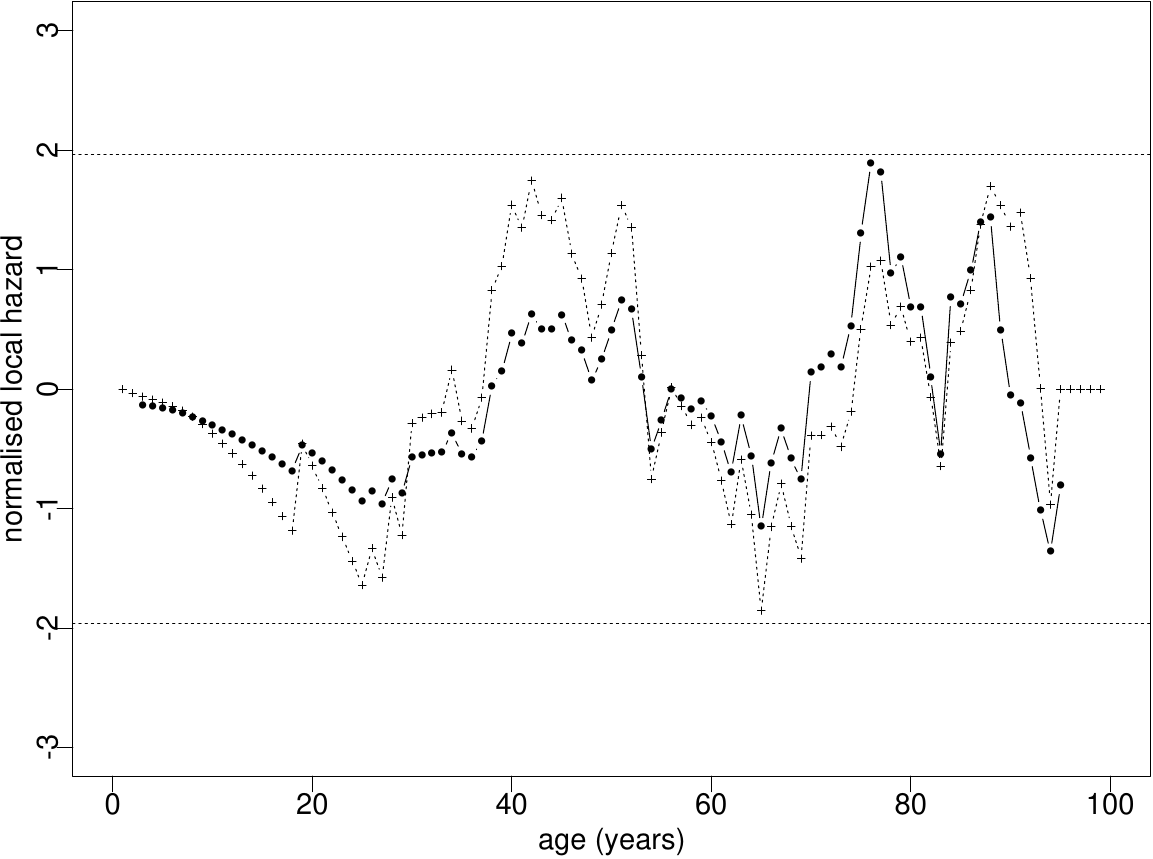}}

\bigskip

\centerline{\includegraphics[scale=0.55]{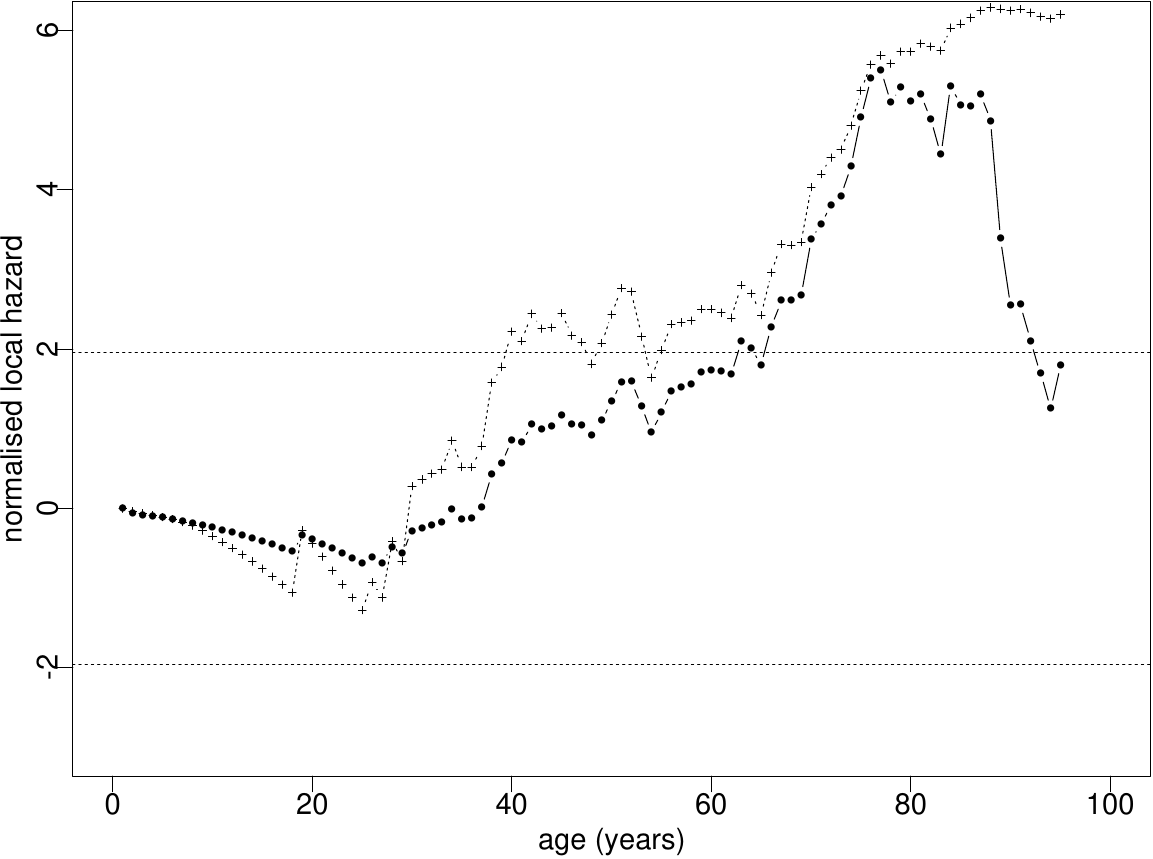}}

{\bigskip\narrower\noindent\sl
{\csc Figure 9.4.} 
Time-discrete NLH plots of Type A (connected dots) and Type B 
(connected plusses) for the Gompertz model are shown for
the group of 716 women in 9.4a and for the 783 men in 9.4b. 
In both cases there is close agreement with the model. 
NLH plots to compare the mortality rate for men with diabetes 
with the estimated Gompertz mortality rate for women with diabetes 
are given in 9.4c, and shows that the mortality rate for men is higher.
\smallskip}

\vfill\eject


\centerline{\includegraphics[scale=0.55]{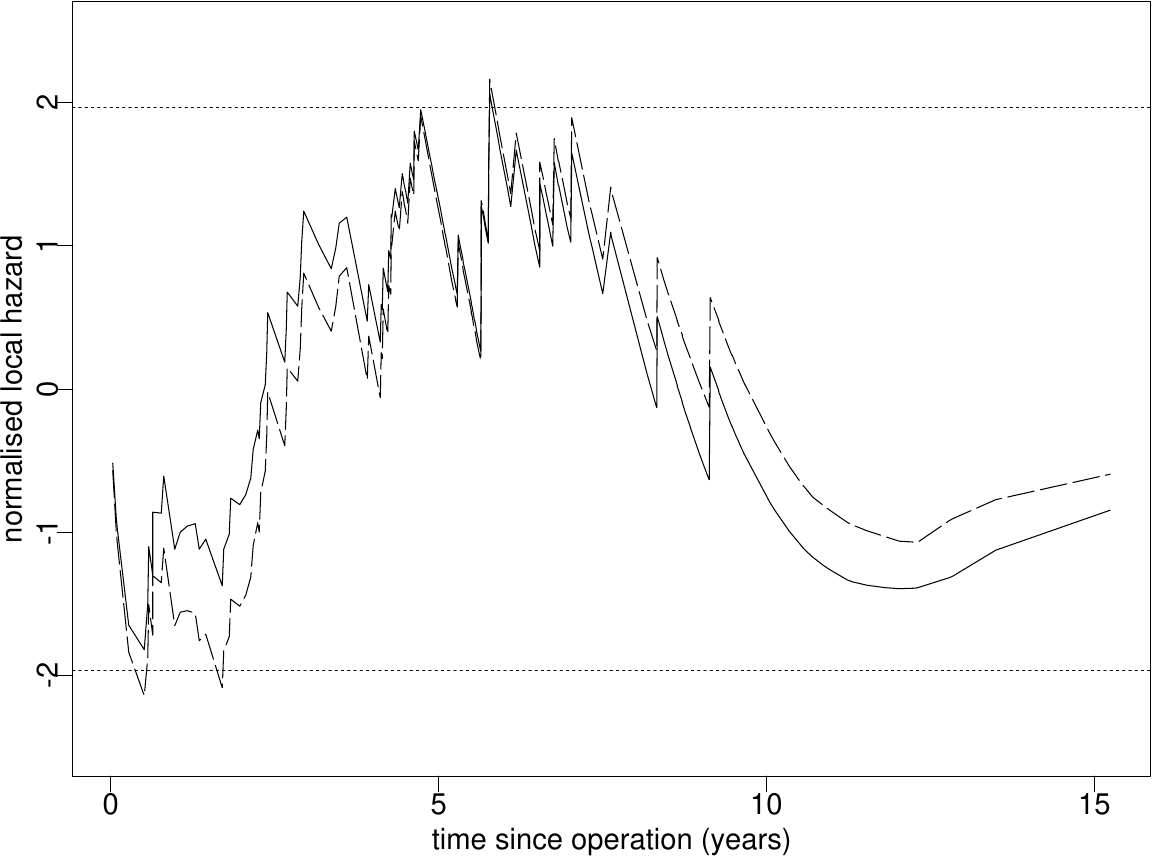}}

\bigskip

\centerline{\includegraphics[scale=0.55]{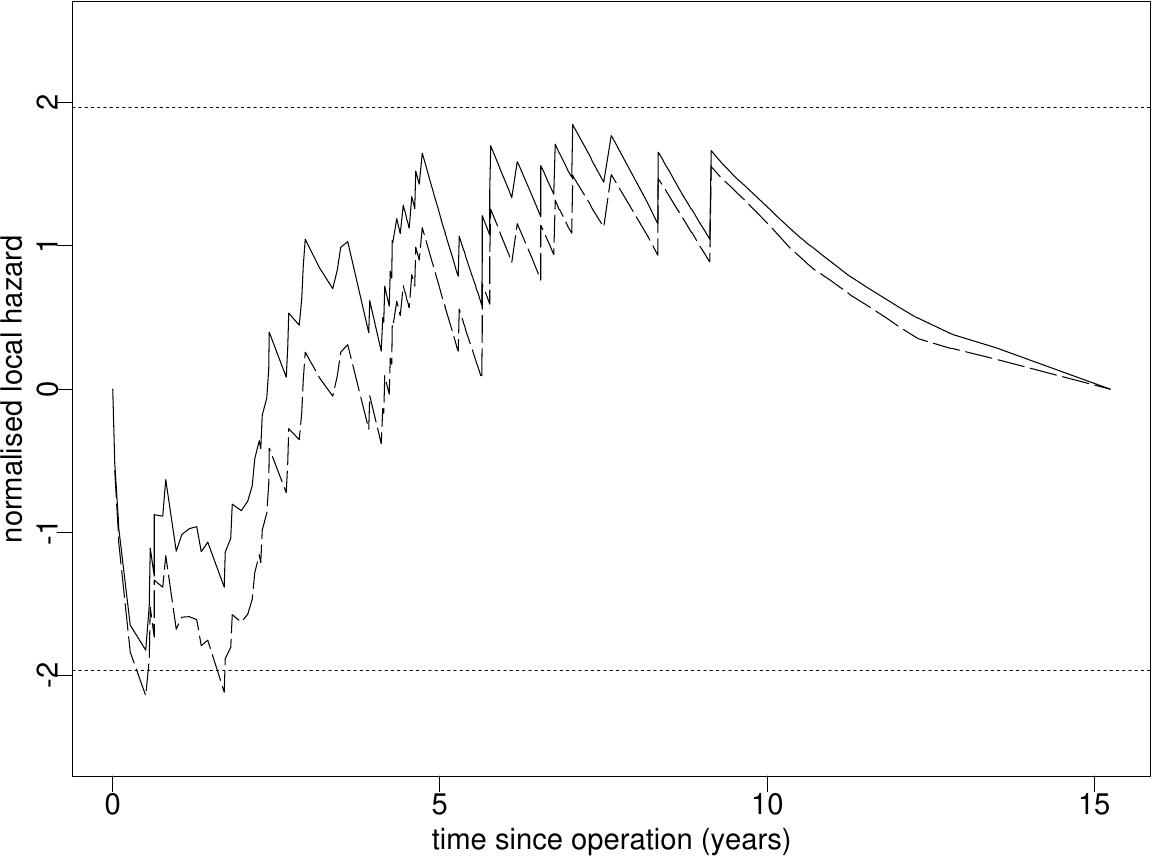}}

\bigskip

\centerline{\includegraphics[scale=0.55]{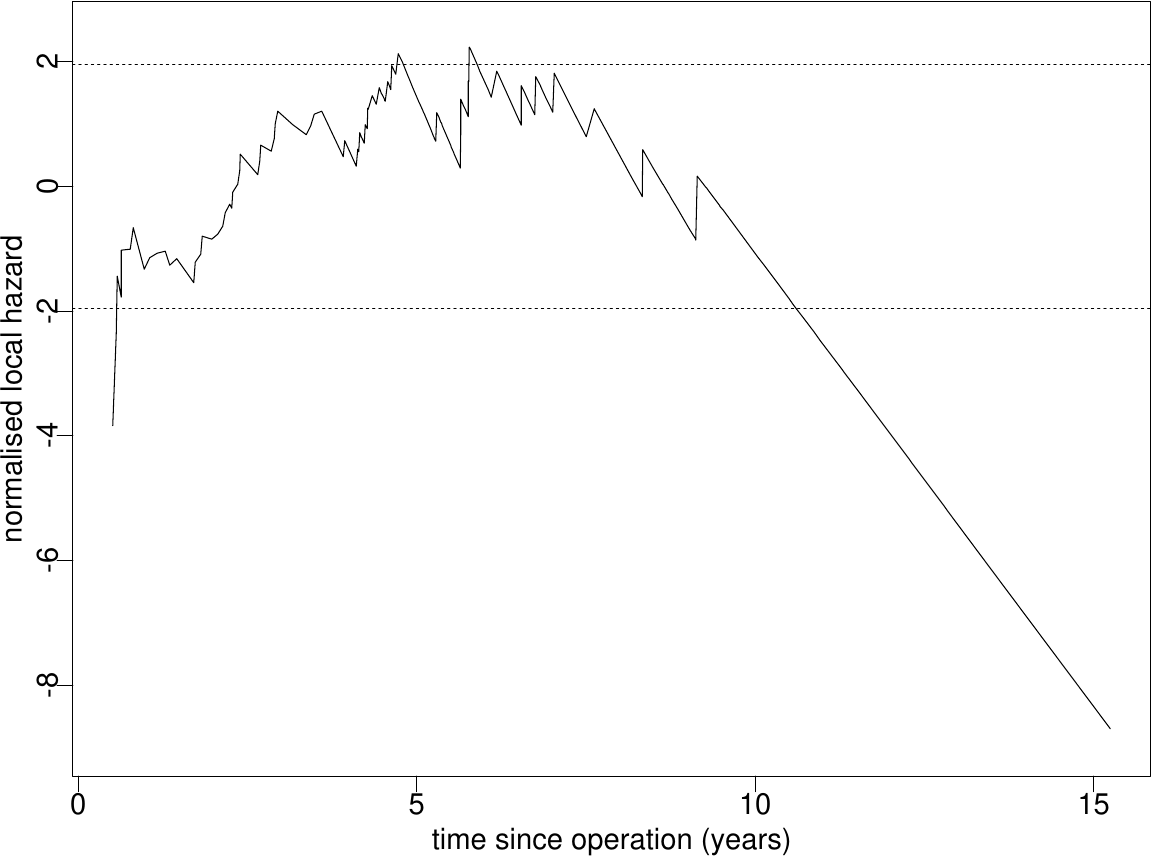}}

\bigskip

\centerline{\includegraphics[scale=0.55]{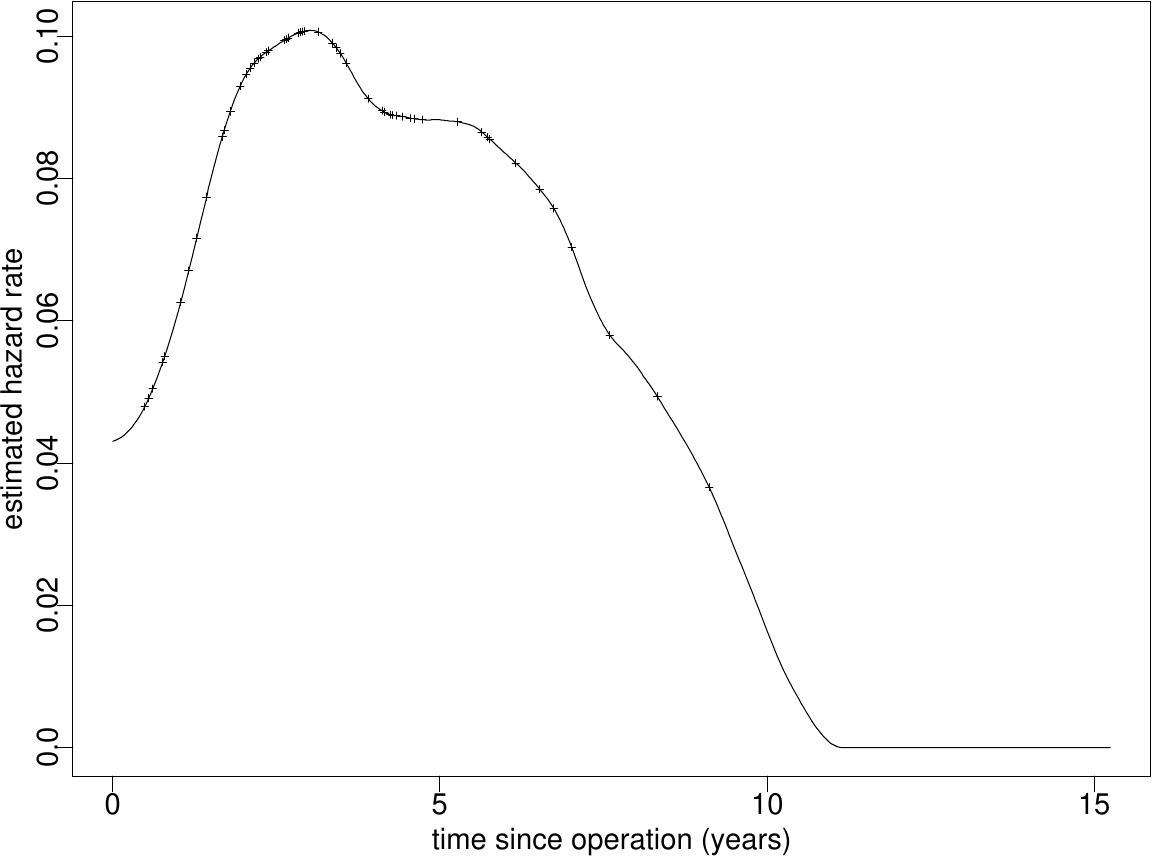}}

{\bigskip\narrower\noindent\sl
{\csc Figure 9.5.} 
NLH-plots of Type A and B are in respectively 9.5a and 9.5b 
for the parametric Cox regression model (dashed lines) 
and the homogeneous exponential model (solid line). 
In 9.5c an A plot for the homogeneous exponential model is given,
suggesting that the true hazard is lower at the beginning and at the end
of the time interval. 
Figure 9.5d gives a kernel smoothed estimate of the 
baseline hazard in a semiparametric Cox model for the melanoma data.  
\smallskip}

\vfill\eject 


\centerline{\includegraphics[scale=0.55]{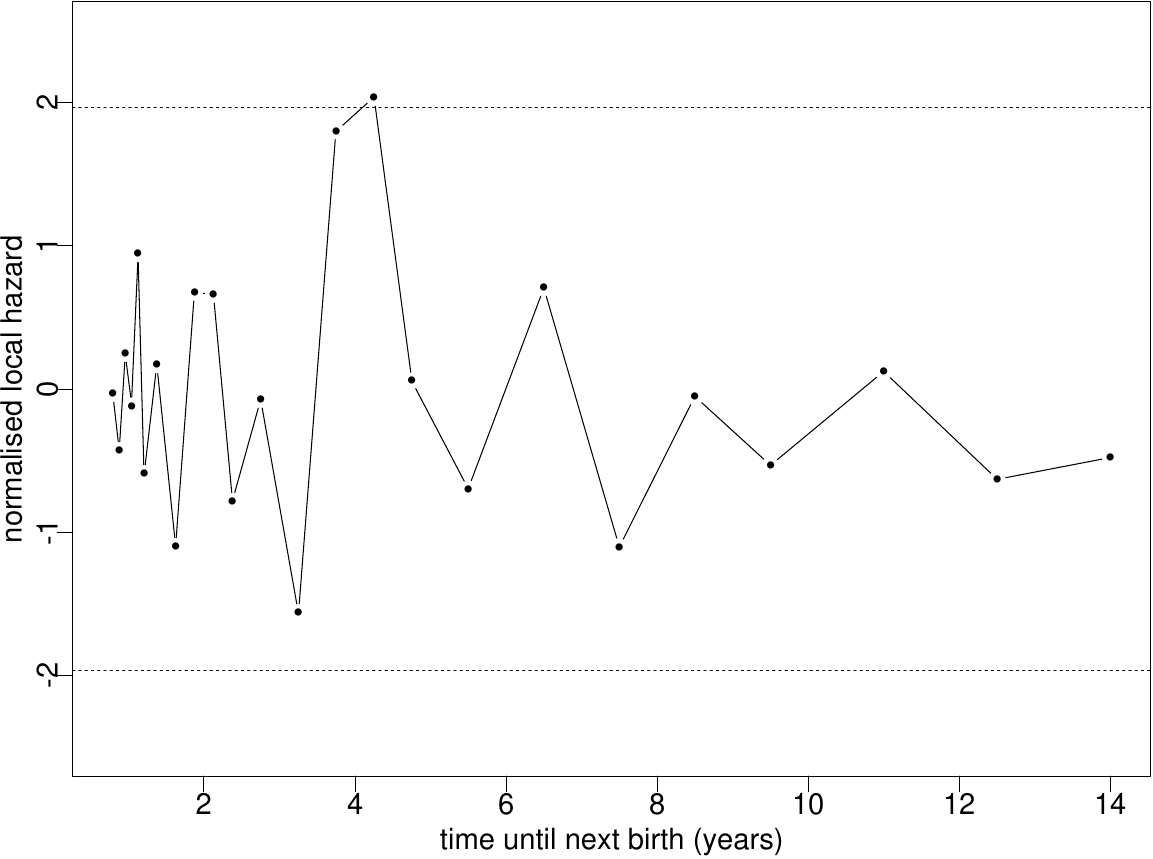}}

{\bigskip\narrower\noindent\sl
{\csc Figure 9.6.} 
Time-discrete normalised hazard difference plot assessing the 
compound Poisson heterogeneity model for the time to next birth following
stillbirth for 451 young Norwegian women. 
\smallskip}

\bye